\documentclass{IEEEtran}
\IEEEoverridecommandlockouts
\usepackage{amsmath,amssymb,amsfonts}
\usepackage{algorithmic, algorithm}
\usepackage{cite, amsfonts, amssymb, amsthm, bm, bbm, graphicx, relsize, multirow, booktabs, tikz,soul}
\usepackage{subcaption}
\usepackage{tabulary}
\usepackage{mathbbol}
\usepackage{textcomp}
\usepackage{xcolor}
\def\BibTeX{{\rm B\kern-.05em{\sc i\kern-.025em b}\kern-.08em T\kern-.1667em\lower.7ex\hbox{E}\kern-.125emX}}

\newcommand{\ds}{\displaystyle}
\newcommand{\norm}[1]{\left\lVert#1\right\rVert}
\newcommand{\bH}{\mathbf{H}}
\newcommand{\ba}{\mathbf{a}}
\newcommand{\bx}{\mathbf{x}}
\newcommand{\wt}{\widetilde}
\newcommand{\bW}{\mathbf{W}}
\newcommand{\bu}{\mathbf{u}}

\newcommand{\br}{\mathbf{r}}
\newcommand{\bz}{\mathbf{z}}
\newcommand{\bv}{\mathbf{v}}
\renewcommand{\i}{\mathrm{i}}

\def\BibTeX{{\rm B\kern-.05em{\sc i\kern-.025em b}\kern-.08em
    T\kern-.1667em\lower.7ex\hbox{E}\kern-.125emX}}
\begin{document}
	This paper  was submitted to the IEEE Transactions on Wireless Communications on September 1, 2021. It has been accepted for publication on April 25, 2022. 
	
	\bigskip
	
	\copyright 2022 IEEE. Personal use of this material is permitted. Permission from IEEE must be obtained for all other uses, in any current or future media, including reprinting/republishing this material for advertising or promotional purposes, creating new collective works, for resale or redistribution to servers or lists, or reuse of any copyrighted  component of this work in other works.
	
	\newpage

	\bstctlcite{IEEE_nodash:BSTcontrol}

\title{Multi-UE Multi-AP Beam Alignment in User-Centric Cell-Free Massive MIMO Systems Operating at mmWave}
\author{Stefano Buzzi,~\IEEEmembership{ Senior Member,~IEEE,}
	Carmen D'Andrea,~\IEEEmembership{ Member,~IEEE,}  Maria Fresia and Xiaofeng Wu
	\thanks{This work was supported by HiSilicon through cooperation agreement YBN2018115022.}
	\thanks{This paper has been partly presented at \textit{IEEE 2021 Globecom} (December 2021) and at the \textit{2021 ITG Workshop on Smart Antennas} (November 2021).}
	\thanks{S. Buzzi and C. D'Andrea are with the Department
		of Electrical and Information Engineering, University of Cassino and Southern Latium, 03043, Cassino,
		Italy, and with Consorzio Nazionale Interuniversitario per le Telecomunicazioni (CNIT), 43124, Parma, Italy.
		M. Fresia and X. Wu are with Huawei Technologies Duesseldorf GmbH Wireless Terminal Chipset Technology Lab, Munich, Germany.}}

\maketitle

\begin{abstract}
This paper considers the problem of beam alignment in a cell-free massive MIMO deployment with multiple access points (APs) and multiple user equipments (UEs) simultaneously operating in the same millimeter wave frequency band.  
Assuming the availability of a control channel at sub-6 GHz frequencies, a protocol is developed that permits estimating, for each UE, the strongest propagation path from each of the surrounding APs, and to perform user-centric association between the UEs and the APs. Estimation of the strongest paths from nearby APs is realized at the UE in a one-phase procedure, during which all the APs simultaneously transmit on pseudo-randomly selected channels with pseudo-random transmit beamformers. An algorithm for orthogonal channels assignment to the APs is also proposed, with the aim of minimizing the mutual interference between APs that transmit on the same channels. 
The performance of the proposed strategy is evaluated both in terms of probability of correct detection of the directions of arrival and of departure associated to the strongest beam from nearby APs, and in terms of downlink and uplink signal-to-interference-plus-noise ratio. Numerical results show that the proposed approach is effective and capable of efficiently realizing beam alignment in a multi-UE multi-AP wireless scenario. 
\end{abstract}


\section{Introduction}
The use of millimeter wave (mmWave) carrier frequencies is one of the main pillars of current and future cellular wireless system \cite{whatwill5Gbe,tripathi2021millimeter}. Indeed, the large bandwidths available at mmWave can greatly contribute to provide the huge data rates that are requested to implement current and future ultra-broadband mobile services. In cellular generations prior to 5G, mmWaves had not been considered due to the increased path loss with respect to the sub-6GHz frequencies, which made them clearly unsuited for communication over distances typical of wireless cellular systems.  Additionally, propagation at mmWave happens only through direct links and/or one-hop reflections, and this poses another further challenge for their use with respect to sub-6 GHz frequencies, where instead multiple reflections and diffraction take place.  Despite these difficulties, in recent years mmWave carrier frequencies have been considered for adoption in wireless cellular communications mainly for two reasons: (a) the use of radio cells of small size has made shorter the typical distance of a wireless cellular link, thus implying that the increased path-loss introduced at mmWave carrier frequencies may become manageable; and (b) the development of multiple antenna communications has led to antennas with an effective aperture that can be practically independent of the wavelength, thus potentially overcoming the path-loss limitation. 
However, in order to be able to really overcome the large path loss and make mmWave links work reliably, narrow beams must be used so as to concentrate radiated energy along those spatial directions associated with the existing line of sight path and/or one hop reflected paths. Otherwise stated, while at sub-6GHz frequencies communication may get started even with fairly broad beams, at mmWave narrow beams are to be used along some channel and geometry dependent directions. The problem of finding those beamforming directions is called \textit{beam alignment} (BA). More precisely, BA is a task that must be accomplished at mmWave in order to ensure that an active link with a sufficiently high signal strength can be established between the intended transmitter and the intended receiver. It is a necessary task that must be executed \textit{before} actual data communication takes place. 

The fifth generation of wireless networks has also seen the introduction of the so-called massive MIMO technology \cite{bjornson2017massive}, which enables the simultaneous transmission to several UEs using the same time-frequency slot, and with a minimum amount of interference. Although providing excellent multiplexing gains, unfortunately this technology does not solve the “cell-edge” problem: when a UE is located in between the reference base station (BS) and an interfering BS, it experiences a poor signal-to-interference-plus-noise ratio (SINR), and, thus, interference management algorithms are to be run, which eventually decrease the spectral efficiency. One solution aimed at overcoming this problem is the so-called cell-free massive MIMO (CF-mMIMO) network deployment, where the macro-BSs are substituted by several APs, having a lower number of antennas and lower complexity \cite{ngo2017cell,buzzi2017cell,interdonato2019ubiquitous}. The APs are assumed to be connected to a central processing unit (CPU) through some wired or wireless connection, and can jointly serve the UEs using the same time-frequency slot. A wise association between the APs and the UEs can be realized, letting each UE be served by a certain number of APs, typically the ones that are closer to the UE of interest or the ones with the highest large scale fading coefficient. This latter deployment is called also “user-centric”, since the set of APs serving a particular UE forms a cluster with the UE at its center. CF-mMIMO user-centric deployments permit to alleviate the aforementioned cell-edge problem, since, given the large number of distributed APs, there is large likelihood that each UE happens to be located very close to at least one AP, which ensures thus a good SINR and a reliable connection. CF-mMIMO architectures are currently widely investigated and are credited to be one of the key network architectures for beyond-5G wireless networks \cite{zhang2019cell}.

Most of the research on CF-mMIMO has been so far mainly carried with reference to sub-6 GHz frequencies, while fewer studies (such as, for instance, \cite{alonzo2019energy,femenias2019cell,guo2021robust}) have addressed CF-mMIMO systems operating at mmWave carrier frequencies. Nonetheless, it is anticipated that in crowded areas with high-demand for mobile broadband services both a large number of distributed antenna  (such as CF-mMIMO) and high-carrier frequencies will be needed, thus implying that CF-mMIMO at mmWave will be one of the typical deployments for future beyond-5G and 6G scenarios. This paper investigates the problem of BA in a scenario wherein multiple APs and multiple UEs use the same frequency band. As illustrated in the following review of the state of the art, the vast majority of existing papers on BA algorithms assume single-user settings wherein a single UE and a single AP (or BS) have to align their beams: applying such procedures in a multi-UE, multi-AP setting would results in a lengthy and non-feasible procedure where each AP-UE pair should align the beams with the remaining devices being silent. In this paper, instead, a procedure for simultaneous BA for multiple UEs and multiple APs is proposed and analyzed.

\subsection{Previous contributions}	
The problem of BA for wireless networks operating at mmWave frequencies has received considerable attention in the recent past. 
{Starting from scenarios with single transmitter and single receiver, paper  \cite{song2015adaptive}
considers the problem of BA using dual-polarized antennas, so that orthogonal polarizations can be sounded in parallel, and proposes a soft-decision algorithm under the assumptions of Ricean-distributed channels with large Ricean $K$ factor and poor scattering environment.  In
\cite{hassanieh2018fast} the authors propose a protocol for fast BA and, instead of sequentially scanning the space with narrow beams, exploit multi-finger beams so as to reduce the time needed to perform BA. A similar approach is also taken in  \cite{Caire_scalable_robust_BA_TCOM2018,song2019efficient}; these papers focus on a system using multi-carrier and single-carrier modulations, respectively, and propose to use compressive sensing to perform BA using transmit and receive pseudo-random multi-finger beamformers. 
Additionally, paper  \cite{liu2017millimeter} develops a theoretical performance analysis of the BA process for both exhaustive search and hierarchical search. The asymptotic expression (in the limit of large length of the training period) of the misalignment probability is derived for both mentioned search scheme and under the assumption that both the UE and the BS have a finite cardinality beamforming codebook. In \cite{maschietti2017robust}, authors consider the problem of robust BA based on the knowledge of the location of the AP and the UE. Since this information can be affected by estimation error, a robust algorithm based on Bayesian team decision is proposed. 
Reference \cite{Chavva_TWC2021} focuses on the problem of beam selection taking into account the random changes of the user orientation deriving closed form expression exploiting statistical models characterizing the time-evolution of non-stationary beam.}
Paper 
\cite{hussain2017throughput}
considers the optimization of the share of time devoted to BA and the share of time dedicated to actual data communication, with the aim of maximizing the system throughput, and derives the optimal beam search parameters adopting the framework of Markov decision processes. The analysis reveals that the BA bisection search algorithm achieves better performance than BA iterative and exhaustive search algorithms. {Reference \cite{Lei_JSTSP2022} adopts the Markov decision process framework to solve the problem of the joint beam training and data communication, through the use of reinforcement learning, in a single-user scenario.}
In \cite{li2019fast} authors formulate the BA problem as a sparse encoding and phaseless decoding problem, and the proposed algorithm can perfectly recover the support and magnitude of the sparse signal (uniquely associated to the beams' directions) in the noiseless case.
In 
\cite{li2019explore}, instead, 
a two-stage procedure is proposed for BA. In the first stage, the algorithm explores and trains all candidate beam pairs, and, then, eliminates a set of less favourable pairs learned from the received signal profile. In the second stage, the algorithm takes an extra measurement for the each of the survived pairs and combines with the previous measurement to determine the best one.
{ In \cite{va2016beam}, the problem of beam switching is considered in high-mobility scenarios: beam switching happens when BA has been already accomplished, and, due to UE mobility and/or to changes in the surrounding scenario, a beam switch is required to ensure continuity of the communication. Machine learning performing BA is considered in reference \cite{rezaie2020location}, again for a point-to-point link; the paper proposes a technique leveraging knowledge of position and orientation of the transmit and receive antenna arrays. Reference \cite{Gao_TCOM_2021} designs a deep neural network, named FusionNet, aimed at beam prediction exploiting the use of an auxiliary sub-6 GHz channel. }

{The above references, despite good performance, focus however on a point-to-point single link. Regarding multiuser scenarios with single BS, reference \cite{Choi_TCOM_2015} proposes a low-complexity beam selection method based on compressive sensing exploiting the mmWave channel sparsity. Paper \cite{Pal_COMML_2018} develops a beam selection algorithm for estimating the best beam for each user in the system. The method attempts to	maximize the sum-rate and nulls-out the multiuser interference. The authors of \cite{Li_WCL_2021} propose a joint multi-beam and channel tracking scheme to perform the beam management in mmWave multiuser systems, optimizing the pilot allocation and using a multicarrier modulation format. Reference \cite{Zhang_TCOM_2021} develops a training beam sequence design for multiuser mmWave tracking systems. The authors formulate a nonlinear optimization problem aimed at the minimization of the average mean squares error of the estimated beams. Reference \cite{Li_WCL_2021} focuses on joint real-time phase shifter network calibration and beam tracking with single-antenna users in the presence of user mobility. In each tracking period, the exploration strategy is determined according to the previously tracked beam directions. Then, based on the latest observation and the historical observations, the time-varying channels and the time-varying phase shifter network deviations are jointly tracked and updated. 
Papers \cite{khojastepour2020multi,ma2020machine, Hu_JSAC_2021} tackle the problem of BA and beam selection in a multiuser scenario applying machine learning tools to solve the problem. 
In the above papers, only one BS is considered, thus the proposed procedures are not applicable in a scenario with multiple APs and multiple users as in a CF-mMIMO deployment.}

\subsection{Paper contribution}
{
This paper, to the best of our knowledge, is the first to consider the problem of BA in a multi-UE multi-AP environment, proposing a BA procedure wherein all the UEs are capable of simultaneously estimating the angle of departure (AoD) and angle of arrival (AoA) of the strongest beam coming from the APs in the neighbours. 
The content of this paper thus enables cell-free coordinated multi-point transmission in a wireless network operating at mmWave and with several distributed APs. 
  
Specifically, the contribution of this work may be summarized as follows. 
A methodology for performing \textit{simultaneous} BA in a multi-AP, multi-UE setup, wherein all the APs and all the UEs operate using the same frequency band, is developed. The methodology consists of a protocol involving the CPU, the UEs, the APs and a macro-BS managing a reliable control channel at sub-6GHz frequency. 
The proposed BA procedure is based on a one-way transmission from the APs, with the UEs operating in listening mode and estimating the AoD and AoA of the strongest beam from the surrounding APs. This is in sharp contrast with the majority of existing BA procedures, that rely on several steps where the UE and the AP iteratively refine their beams.  
In order to make the UEs capable to discriminate the signals coming from different APs, these must transmit on disjoint set of orthogonal channels. Since, in general, the number of APs is much greater than the number of available sets of orthogonal channels, a further contribution of this paper is an algorithm aimed at assigning the sets to the APs, with the objective of minimizing the mutual interference among APs using the same set of orthogonal carriers.
Two different BA algorithms, to be implemented at the UE,  are proposed, one inspired from \cite{Caire_scalable_robust_BA_TCOM2018}, and one totally original.
{ The latter algorithm is also generalized to the case in which beam tracking is to be performed in order to cope with user mobility.}
The study is finally completed by a thorough simulation-based performance study,  in terms of probability of correct detection of the couple (AoA, AoD) of the strongest beams, and in terms of achievable spectral efficiency. The obtained results will show the effectiveness of the proposed procedure.}

{Part of the technical material here presented has appeared in a preliminary and reduced-length format in the conference papers \cite{Buzzi_BA_WSA2021,Buzzi_BA_Globecom2021}. This manuscript contains a more organized and refined presentation of the topic and of the developed algorithm, as well as new material such as the extension of the BA procedures in a dynamic environment and the consideration of the impact of the proposed BA algorithm on the performance of the subsequent data transmission phase. Moreover, the majority of the shown numerical results are also original.
}

\subsection{Notation}
In the following, lower-case and upper-case non-bold letters are used for scalars, $a$ and $A$, lower-case boldface letters, $\mathbf{a}$, for vectors and upper-case boldface letters, $\mathbf{A}$, for matrices. The transpose, the inverse, the conjugate and the conjugate transpose of a matrix $\mathbf{A}$ are denoted as $\mathbf{A}^T$, $\mathbf{A}^{-1}$, $\mathbf{A}^*$ and $\mathbf{A}^H$, respectively. The $i$-th row  and the $j$-th column of the matrix $\mathbf{A}$ are denoted as $\left(\mathbf{A}\right)_{(i,:)}$ and $\left(\mathbf{A}\right)_{(:,j)}$, respectively. The norm of a vector $\mathbf{a}$ is denoted as $\norm{\mathbf{a}}$. The $N$-dimensional identity matrix is denoted as $\mathbf{I}_N$, the $N\times M$ matrix with all ones is $\mathbf{1}_{N \times M}$ and the the $N\times M$ matrix with all zeros is $\mathbf{0}_{N \times M}$. The complex circularly symmetric Gaussian random variable with mean $\mu$ and variance $\sigma^2$ is denoted as $\mathcal{CN}\left(\mu,\sigma^2\right)$. The set of the complex $N$-dimensional vectors is denoted as $\mathbb{C}^{N}$ and $\i$ is the imaginary unit.

\begin{small}
\begin{figure}
	\begin{center}
		\includegraphics[scale=0.3]{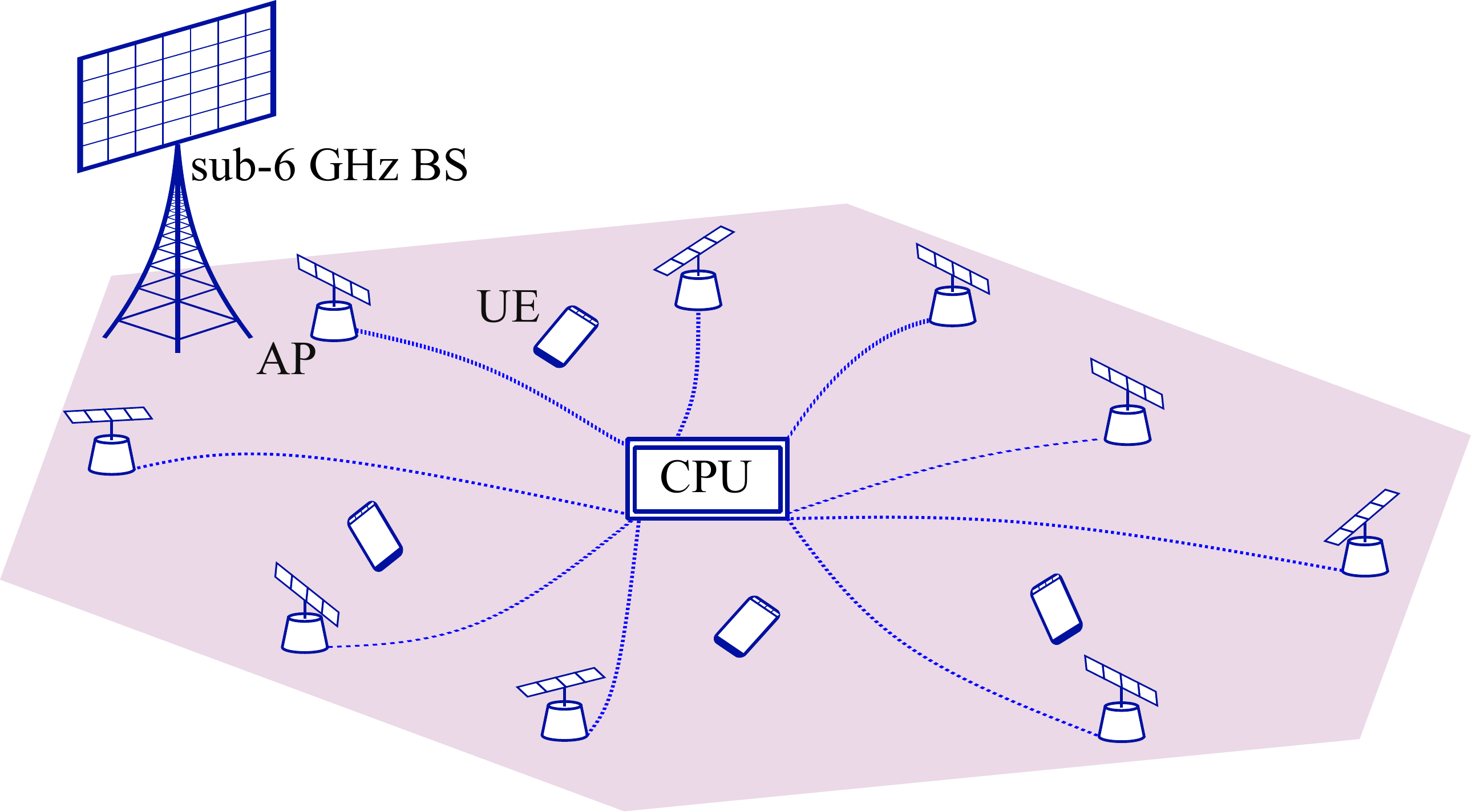}
	\end{center}
	\caption{{Considered scenario.}}
	\label{Fig:scenario}
\end{figure} 
\end{small}

\section{System description}
Consider a CF-mMIMO system where $M$ APs simultaneously serve $K$ UEs on a shared channel. We focus on the BA procedure and, for the sake of simplicity, consider a bi-dimensional layout\footnote{Extension to 3D layouts can be straightforwardly done.}. 
We use the following notation:
\begin{itemize}
	\item[-] $N_{\rm UE}$ denotes the number of antennas at the generic UE. 
	\item[-] $N_{\rm AP}$ denotes the number of antennas at the generic AP. 
	\item[-] $n_{\rm UE}< N_{\rm UE}$ denotes the number of RF chains at the generic UE. 
	\item[-] $n_{\rm AP}< N_{\rm AP}$ denotes the number of RF chains at the generic AP. 
\end{itemize} 
The number of antennas and RF chains is taken constant for all the APs and all the UEs to simplify notation, but extension to the general case where each AP and UE has an arbitrary number of antennas and RF chains is straightforward.
Both the APs and the UEs are equipped with uniform linear arrays (ULAs) with random orientations, and the steering angles are assumed to take values in the range $[-\pi/2,\pi/2]$. See Fig. \ref{Fig:scenario} for a sample scenario realization.

\begin{figure}
	\begin{center}
	\includegraphics[scale=0.22]{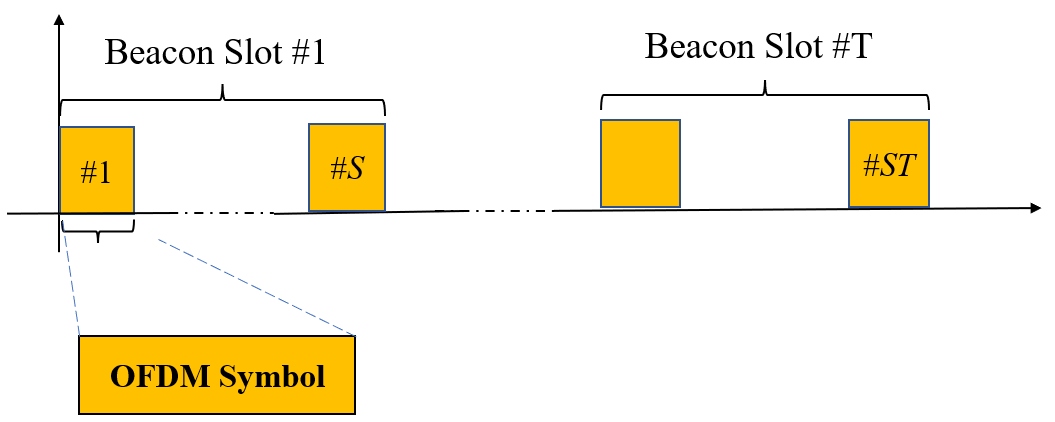}
	\caption{{Considered frame format.}}
	\label{Fig:frame}
	\end{center}
\end{figure}

\subsection{Transmission format}
It is assumed that the adopted modulation format is the orthogonal frequency division multiplexing (OFDM); the OFDM symbol duration is denoted by $t_0$, $B$ denotes the overall available bandwidth, while  the subcarrier spacing for the OFDM signal is denoted by  $\Delta f$. This implies that the number of subcarriers is $N_C=B/\Delta f$. The OFDM symbol duration is  $1/\Delta f + \tau_{CP}$, with $\tau_{CP}$ the length of the cyclic prefix. The BA phase will span $T$ \textit{beacon slots} (a terminology borrowed from \cite{Caire_scalable_robust_BA_TCOM2018}), each made of $S$ consecutive OFDM symbols. See also Fig. \ref{Fig:frame} for a graphical representation of the considered frame format.

\subsection{Channel model}

The downlink channel between the $m$-th AP and the $k$-th UE in the $s$-th beacon slot is represented by an $(N_{\rm UE} \times N_{\rm AP})$-dimensional matrix-valued linear time invariant (LTI) system with impulse response\cite{Heath_SP_mmwave_2016,DAndrea_EUSIPCO2021}
\begin{equation}
\bH_{k,m}^{(s)}(\tau)\!\!=\!\!\ds \sum_{\ell'=0}^{L_{k,m}} \!\!\alpha_{k,m,\ell'}^{(s)}
\ba_{\rm UE}(\varphi_{k,m,\ell'}) \ba_{\rm AP}^H(\theta_{k,m,\ell'}) \delta(\tau-\tau_{k,m,\ell'}) \; .
\label{eq:channelmodel_km} 
\end{equation} 
In the above equation, we have that
\begin{itemize}
	\item[-] $L_{k,m}$ denotes the number of paths that contribute to the channel between the $k$-th UE and the $m$-th AP. This number depends on the geometry of the system. Usually in a poor scattering environment typical of mmWave frequencies we have that $L_{k,m} << \min\{N_{\rm AP}, N_{\rm UE}\}$.
	\item[-] $\alpha_{k,m,\ell'}^{(s)}$ is the complex gain associated to the $\ell'$-th path in the $s$-th beacon slot. We assume $\alpha_{k,m,\ell'}^{(s)} \sim \mathcal{CN}(0, \gamma_{k,m,\ell'})$, with $\gamma_{k,m,\ell'}$ denoting the reflection coefficient variance. 
	\item[-] $\varphi_{k,m,\ell'}$ and $\theta_{k,m,\ell'}$ are the AoA and AoD relative to the $\ell'$-th path.
	\item[-] $\ba_{\rm AP}(\cdot)$ and $\ba_{\rm UE}(\cdot)$ are the ULA array responses at the AP and at the UE, respectively. Assuming half-wavelength spacing for the array elements, they are expressed as
	\[
	\begin{array}{lll}
		\ba_{\rm AP}(\varphi)=[1, \; e^{\i \pi \sin \varphi}, \; \ldots, \; e^{\i \pi (N_{\rm AP}-1)\sin \varphi}]^T \; , \quad \\
		\ba_{\rm UE}(\theta)=[1, \; e^{\i \pi \sin \theta}, \; \ldots, \; e^{\i \pi (N_{\rm UE}-1)\sin \theta}]^T \; , \\
	\end{array}
	\]
	\item[-] $\tau_{k,m,\ell'}$ is the propagation delay associated to the $\ell'$-th path. 
\end{itemize}

Notice that while the complex gains $\alpha_{k,m,\ell'}^{(s)}$ depend on the beacon slot index $s$, this does not happen for the other parameters, such as the number of paths, their associated delays, and the corresponding AoAa and AoDs, which typically vary over much larger timescales than the complex gains associated with propagation paths. Otherwise stated, $\alpha_{k,m,\ell'}^{(s)}$ accounts for the fast small-scale fading, while the remaining channel parameters are tied to large-scale variations.

\section{Beam alignment procedure preliminaries} \label{BA_procedure}

\subsection{The data-patterns}
Before the BA procedure starts, a set of  resources, referred to as  \textit{data-patterns}, are to be defined and assigned to the APs. Obviously, since in a large system the number of  orthogonal data-patterns can be reasonably assumed to be smaller than the number of APs, the same data-pattern is to be reused across the network. We will tackle later this issue. 

Two different types of data-patterns will be considered in this paper. 
With regard to the former type, we define as data-pattern a set of subcarriers and beamforming vectors, which the APs use to transmit \textit{constant} signals. 
Since each AP is equipped with $n_{\rm AP}$ RF chains, i.e., it can simultaneously transmit $n_{\rm AP}$ data streams using different beamforming vectors. In order to permit data stream separation at the UEs without having knowledge of the AP locations and antenna array orientation, it is needed that the transmitted data streams  are orthogonal \textit{before} beamforming.  One way of achieving this is through the use of  non-overlapping subcarriers for the parallel data-streams.
Denoting by $Q$ the number of subcarriers assigned to each AP RF 
chain\footnote{Letting $Q>1$ permits sounding the channel at multiple frequency locations, and is expected to lead to improved BA capability.}, 
one can easily realize that the number of available different data-patterns is  $D= \lfloor \lfloor N_C/Q\rfloor / n_{\rm AP} \rfloor$; the corresponding $D$ data-patterns will be denoted by  ${\cal D}_1, \ldots, {\cal D}_D$. 
The generic set ${\cal D}_d$ will thus specify the $Q$ subcarriers \textit{and} the beamformers to be used in each beacon slot on a certain RF chain of a certain AP. More precisely, letting $\mathcal{L}_{d,s,i}$ and $\mathcal{U}_{d,s,i}$  denote the set of $Q$ subcarriers and the set of transmit beamformers, respectively, to be used by the APs that are assigned the $d$-th data-pattern on the $i$-th RF chain, the data-pattern ${\cal D}_d$ is in this case formally described as
\begin{equation}
	\begin{array}{llll}
		{\cal D}_d=& 
		\left\{ \left\{ \mathcal{L}_{d,s,i}, \, s=1, \ldots, T , \; i=1, \ldots, n_{\rm AP} \right\} ,\,
		\right. \\ & \left.
		  \left\{ \mathcal{U}_{d,s,i}, \, s=1, \ldots, T , \; i=1, \ldots, n_{\rm AP} \right\}
		\right\} \, .
	\end{array}
	\label{eq:Dd}
\end{equation}


Since, as already highlighted, APs are assumed to transmit a constant signal using the above defined data-patterns, we refer to them as \textit{pilot-less} data-patterns. 

The latter definition of data-pattern, by allowing the transmission of modulated signals, leads to a larger number of orthogonal data-patterns, and permits to increase the distance between conflicting APs that will have to be assigned the same data-pattern, eventually resulting in better performance. 
Precisely, in each beacon slot, APs assigned the same pilot-less data-pattern may be differentiated by allowing them to transmit orthogonal pilots of length $S$. 
Since up to $S$ different orthogonal pilots of length $S$ can be generated, this strategy increases to $SD$ the number of available orthogonal data-patterns. 
The $\ell$-th pilot sequence,  $\bm{\phi}_{\ell}$ say, of length $S$ is in particular defined as 
$
	\bm{\phi}_{\ell}=\left[  \sqrt{\beta} e^{\i \widetilde{\phi}_{\ell}(1)}, \ldots,  \sqrt{\beta} e^{\i \widetilde{\phi}_{\ell}(S)} \right]^T
$,
and fulfills the relation $\bm{\phi}_{\ell}^H\bm{\phi}_{\ell'}=S \beta \delta ( \ell-\ell')$,
with $\beta>0$ the power transmitted by the APs in each beacon-slot and on each subcarrier. This latter type of data-pattern will be named   \textit{pilot-based} data-patterns. Clearly, the pilot-less definition can be seen as a particular case of the pilot-based definition where $\widetilde{\phi}_{\ell}(i)=0, \; \forall i=1,\ldots, S, \; \ell=1, \ldots , S$. For this reason, in the following we will describe the BA procedure assuming the more general case of pilot-based data-patterns and will assess the performance difference between the two types of data-patterns in  Section \ref{Numerical_results}.

\subsection{Location-based data-patterns assignment algorithm} \label{LBA_procedure}
We now discuss how  the data-patterns are to be shared among the APs in the general case in which the number of APs is larger than the number of available data-patterns. In general, the distance between the APs using the same set of resources should be as large as possible. We thus propose a location-based (LB) data-pattern assignment  in order to reduce the \textit{BA contamination} in the system. 
{It is thus assumed that AP positions are known to the network operator; otherwise stated, we consider the relevant case of a network deployment with fixed topology, leaving aside the special case of a network with mobile APs.}
The proposed procedure uses the well-known $k$-means clustering method \cite{K_means_steinley2006}, i.e., an iterative algorithm that is able to partition APs into disjoint clusters.
Defining the \textit{centroid} of each cluster as the mean of the positions of the APs in the cluster, 
the algorithm, accepting as input the APs positions and the number of data-patterns that we generically denote as $\widetilde{D}$,  is summarized in Algorithm \ref{LB_resource_set} and operates as follows:

\begin{itemize}

	\item[a.] $\lceil M/\widetilde{D}\rceil$  centroids are chosen so that they approximately form a regular grid over the considered area, {i.e., the so-called parameter "$k$" in the $k$-means clustering is $\lceil M/\widetilde{D}\rceil$.}
	
	\item[b.] Assign  each AP to its nearest centroid, with the constraint that no more than $\widetilde{D}$ APs are assigned to each centroid. This way, the APs are thus divided in clusters of no more than  $\widetilde{D}$ elements.
	
	\item[c.] Update the centroid positions by averaging over the positions of the APs belonging to each cluster.
	
	\item[d.] Repeat steps [b.] and [c] until the positions of the centroids converge. 
	
	\item[e.] Once the AP clusters have been defined, data-patterns are to be assigned to the APs according to the following strategy.  We characterize the APs in each cluster with their position relative to the centroid of the cluster to which they belong. Then, we assign the first data-pattern to the AP in each cluster that has the largest latitude (i.e. the most northern one); the second data-pattern to the AP in each cluster with the second largest latitude, and so on. The assignment procedure stops when all the the APs in the system have been assigned a data-pattern.
\end{itemize}

{Regarding step [e.], its aim is to ensure that APs that are assigned the same data-pattern are not too close. The methodology that we propose here, based on the latitude only of the APs and not on their full 2D coordinates, is clearly suboptimal, but has been tested to represent a good trade-off between complexity and performance.} Finally, note that $\widetilde{D}=D$ when pilot-less data-patterns are used, and $\widetilde{D}=SD$ for the case in which pilot-based data-patterns are adopted.

\begin{small}
\begin{algorithm}
	
	\caption{Location-based data-patterns assignment algorithm}
	
	\begin{algorithmic}[1]
		
		\label{LB_resource_set}
		\STATE  Allocate  $\lceil M/ \widetilde{D} \rceil$ {centroids}
		so as to form an approximately regular grid over the considered area.
		\REPEAT 
		
		\STATE Assign each AP to the nearest centroid with the constraint that no more than $\widetilde{D}$ APs are associated to the same centroid.
		
		\STATE Compute the new positions of the centroids averaging over the positions of the APs belonging to the same cluster.
		
		\UNTIL convergence of the positions of the centroids or maximum number of iterations reached. 
		
		\STATE Assign the first data-pattern to the AP in each cluster that has the largest latitude (i.e. the most northern one); assign the second data-pattern to the AP in each cluster with the second largest latitude. Continue until all the the APs in the system have been assigned a data-pattern.
		
	\end{algorithmic}
	
\end{algorithm}
\end{small}

\begin{small}
\begin{figure*}
	\begin{center}
		\includegraphics[scale=0.32]{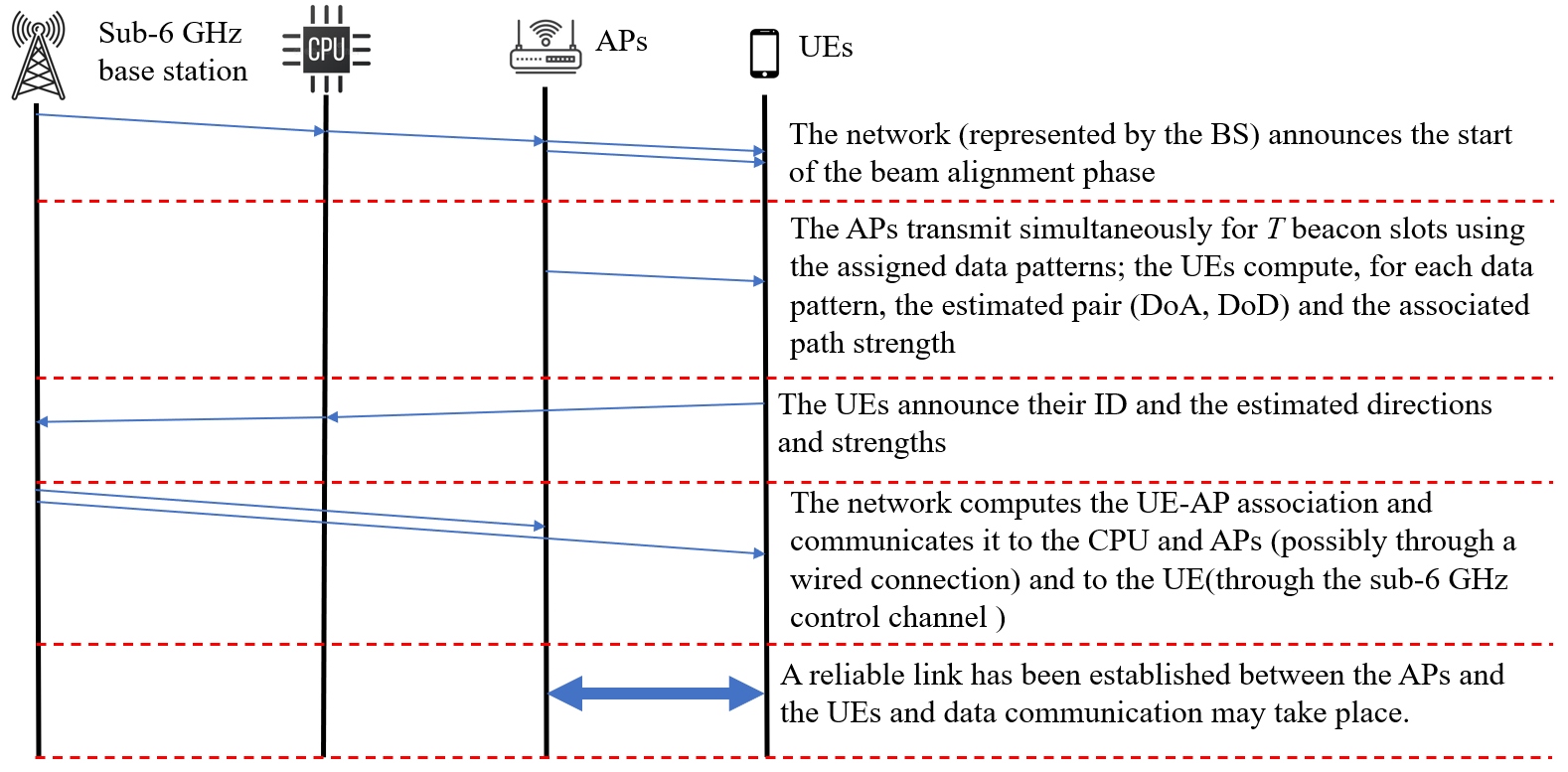}
	\end{center}
\caption{{Temporal diagram of the proposed BA procedure.}}
\label{Fig:time-diagram}
\end{figure*}
\end{small}
\subsection{Timing of the beam alignment procedure}
Similarly to many other papers dealing with the BA problem at mmWave, we assume that a general frame synchronization information is available in the system. This  can be ensured by exploiting the fronthaul connection between the APs and the CPU, and by using a control-plane connection with the UEs at a sub-6 GHz carrier frequency. The BA procedure and the subsequent user association phase is made of the following steps: 
\begin{itemize}
	\item[a)] All the APs transmit simultaneously proper signals on their assigned data-patterns and the UEs gather information and estimate the AoAs and AoDs corresponding to the strongest received paths for each data-pattern. 
	\item[b)] Using the sub-6 GHz uplink control channel, each UE communicates to the network its position and, for each of the 
	 data-patterns\footnote{Remember that the number of data-patterns is either $D$, if the pilot-less definition is used, or $SD$, if the pilot-based definition of data-pattern is used.}, the AoA and AoD of the strongest detected beam and a strength 
	 indicator\footnote{ Notice that the use of the described one-phase  procedure implemented at the UE permits the
	 	realization of the BA simultaneously for an arbitrarily large number of UEs,
	 	and this is one great advantage with respect to the case in which traditional strategies are	used.}.
	\item[c)] Based on the information gathered from all the UEs, the network makes user-centric AP-UE association and communicates AP-UE associations to the APs and to the UEs via the sub-6 GHz control channel.
\end{itemize} 
Refer to Fig. \ref{Fig:time-diagram}  for the temporal diagram of the proposed BA procedure.

\section{Beam Alignment signal model} \label{BA_signal_model}
We are now ready to provide the full details about the signal model.

\subsection{Time-continous model}
Let us focus on the signal transmitted in the $s$-th beacon slot, i.e. for $t \in [st_0, (s+1)t_0]$. The baseband equivalent of the signal transmitted in the $s$-th beacon slot by the $m$-th AP can be expressed through the following $N_{\rm AP}$-dimensional vector-valued waveform:
\begin{equation}
	\bx_{m,s}(t)=\ds \sum_{i=1}^{n_{\rm AP}}x_{m,s,i}(t) \bu_{m,s,i} 	\; ,
\end{equation}
where $x_{m,s,i}(t)$ is the signal corresponding to the $i$-th data stream from the $m$-th AP in the $s$-th beacon interval; the $N_{\rm AP}$-dimensional vector $\bu_{m,s,i}$  is the corresponding transmit beamformer\footnote{Notice that we are here implicitly assuming that the transmit beamformer is kept constant over an entire beacon slot, i.e. for $S$ consecutive OFDM symbols.}. 
The signal received in the $s$-th beacon slot at the $k$-th UE, before the receive beamformer is applied, can be easily shown to be written as
\begin{equation}
	\begin{array}{llll}
		\br_{k,s}(t) = & \ds \sum _{m=1}^M \ds \int \bH_{k,m}^{(s)}(\tau) \bx_{m,s}(t-\tau) d \tau  + \bz_{k,s}(t)= \\ 
		& \ds \sum _{m=1}^M \sum_{\ell'=0}^{L_{k,m}}  \sum_{i=1}^{n_{\rm AP}} \alpha_{k,m,\ell'}^{(s)} g^{(AP)}_{k,m,\ell',s,i} \, \times \\ & x_{m,s,i}(t-\tau_{k,m,\ell'})  \ba_{\rm UE}(\varphi_{k,m,\ell'}) + \bz_{k,s}(t) \; ,
	\end{array}
	\label{eq:received_ks}
\end{equation}
with $g^{(AP)}_{k,m,\ell',s,i}=\ba_{\rm AP}^H(\theta_{k,m,\ell'})\bu_{m,s,i}$ and $\bz_{k,s}(t)$ an $N_{MS}$-dimensional vector waveform representing the AWGN contribution at the $k$-th UE receiver in the $s$-th beacon interval.

The $k$-th UE can apply $n_{\rm UE}$ different receive beamforming vectors to the received signal  \eqref{eq:received_ks}. Denoting by $\bv_{k,s,j}$ the $j$-th beamformer (with $j=1, \ldots, n_{\rm UE}$) used by the $k$-th UE in the $s$-th beacon slot, the following set of observables is available at the $k$-th UE after beamforming:
\begin{equation}
	\begin{array}{llll}
		y_{k,s,j}(t)= &\ds \frac{1}{\sqrt{n_{\rm UE}}} \bv^H_{k,s,j}\br_{k,s}(t) = \\ &
		\ds \sum _{m=1}^M \sum_{\ell'=0}^{L_{k,m}}\sum_{i=1}^{n_{\rm AP}} \ds \frac{1}{\sqrt{n_{\rm UE}}}\alpha_{k,m,\ell'}^{(s)}
		{g^{(AP)}_{k,m,\ell',s,i}} g^{(UE)}_{k, \ell', s, j} \times \\ & \;
		x_{m,s,i}(t-\tau_{k,m,\ell'}) + z_{k,s,j}(t) \; ,
	\end{array}
	\label{eq:received_ksj}
\end{equation}
for $j=1, \ldots, n_{\rm UE}$, with $g^{(UE)}_{k, \ell', s, j}= \bv^H_{k,s,j}\ba_{\rm UE}(\varphi_{k,m,\ell'})$ and $z_{k,s,j}(t)= \frac{1}{\sqrt{n_{\rm UE}}}\bv^H_{k,s,j} \bz_{k,s}(t)$.
The waveforms $y_{k,s,j}(t)$, for all $j$, undergo the usual OFDM receiver processing, and every OFDM symbol in $y_{k,s,j}(t)$ is converted into an $N_C$-dimensional vector. Focusing on the generic $p$-th OFDM symbol, and letting $s(p)=\lfloor p/S \rfloor$ denote the  beacon slot index associated with the $p$-th OFDM symbol, the A/D conversion leads to the $N_C$ scalar entries 
 $Y_{k,p,j}(0), \ldots, Y_{k,p,j}(N_C-1)$.
In particular, it is easy to see that the $q$-th entry of such vector, corresponding to the discrete-time sample on the $q$-th subcarrier,  is expressed as
\begin{equation}
	\begin{array}{llll}
		Y_{k,p,j,i}(q)=&  \ds \frac{1}{\sqrt{n_{\rm UE}}} \ds  \sum _{m=1}^M \bv^H_{k,s(p),j}  
		\mathcal{H}_{k,m}^{(s)}(q) \times \\ & \;
		  X_{m,p,i}(q) \bu_{m,s(p),i} + Z_{k,p,j,i}(q) \; ,
	\end{array}
	\label{eq:Yskj}
\end{equation}
where $X_{m,p,i}(q)$ is the $q$-th data symbol transmitted in the $p$-th OFDM slot on the $i$-th transmit RF chain, $Z_{k,p,j,i}(q)$ contains the AWGN contribution and $\mathcal{H}_{k,m}^{(s)}(q)$ is the matrix-valued Fourier transform of the channel impulse response $\bH_{k,m}^{(s)}(\tau)$ computed at the frequency $q/(t_0)$, i.e.,

\begin{equation*}
	\mathcal{H}_{k,m}^{(s)}(q)\!\!=\!\!\sum_{\ell'=0}^{L_{k,m}} \!\! \alpha_{k,m,\ell'}^{(s)} \ba_{\rm UE}(\varphi_{k,m,\ell'}) \ba_{\rm AP}^H(\theta_{k,m,\ell'})
	e^{-\i 2\pi \frac{q}{t_0}\tau_{k,m,\ell'}}\, .
\end{equation*}
As already said, during the BA phase with pilot-based data-patterns, each AP transmits a pilot sequence that allows to distinguish APs using the same data-pattern. Otherwise stated, the $m$-th AP transmits $X_{m,p,i}(q)=\sqrt{\beta}e^{\i \widetilde{\phi}_{\ell(m)}(p \mod s)}$, where $\ell(m)$ is the index of the pilot sequence assigned to the $m$-th AP, on its assigned subcarriers for $T$ consecutive beacon slots. Assuming that $m$-th AP uses the data-pattern ${\cal D}_d$, this implies that $ q \in \mathcal{L}_{d,s,i}$, with $s=1, \ldots, T$. 

Now, in order to perform direction estimation of the strongest beams from nearby APs, each UE can rely on the knowledge of the data-patterns ${\cal D}_1, \ldots, {\cal D}_D$, and of the orthogonal pilot sequences $\bm{\phi}_{1}, \ldots, \bm{\phi}_{S}$. Based on this information, it has to determine the AoA and AoD of the strongest multipath components to be used for data communication. Notice that no information on the APs location or on the network topology is needed at the UE. The UE will simply determine the strongest directions for the data sensed on each of the system defined data-patterns.

\subsection{Angles discretization and pseudo-random beamforming codebooks}
The AoAs and AoDs,  $\varphi_{k,m,\ell}$ and $\theta_{k,m,\ell}$ in Eq. \eqref{eq:channelmodel_km}, respectively, take continuous values, but in the BA procedure we use the approximate finite-dimensional (discrete) beamspace representation  \cite{Heath_SP_mmwave_2016}. We thus consider the discrete set of AoDs and AoAs
\begin{equation}
	\begin{array}{llll}
		&\Theta=\left\lbrace \widehat{\theta} : \ds \frac{1 + \sin( \widehat{\theta} ) }{2} = \ds \frac{u-1}{N_{ \rm AP}}, \; u=1,\ldots, N_{\rm AP} \right \rbrace \, , \\
		&\Phi=\left\lbrace \widehat{\varphi} : \ds \frac{1 + \sin( \widehat{\varphi} ) }{2} = \ds \frac{u'-1}{N_{\rm UE}}, \; u'=1,\ldots, N_{\rm UE} \right \rbrace
	\end{array}
\label{Discrete_Sets}
\end{equation}
and use the corresponding array responses $\mathcal{F}_{\rm AP}= \left \lbrace  \ba_{\rm AP} ( \widehat{\theta}) : \widehat{\theta} \in \Theta \right \rbrace $ and $\mathcal{F}_{\rm UE}= \left \lbrace  \ba_{\rm UE} ( \widehat{\varphi}) : \widehat{\varphi} \in \Phi \right \rbrace $ as a discrete dictionary to represent the channel response. For the ULAs considered in this approach the dictionaries $\mathcal{F}_{\rm AP}$ and $\mathcal{F}_{\rm UE}$, after suitable normalization, yield orthonormal bases corresponding to the columns of the unitary discrete Fourier transform (DFT) matrices $\bW_{N_{\rm AP}}$ and $\bW_{N_{\rm UE}}$ defined as 
$
		\left[\bW_{N}\right]_{p,p'}=\ds \frac{1}{\sqrt{N}}e^{ \i 2 \pi (p-1)\left(\frac{p'-1}{N} -\frac{1}{2}\right)}$, with $ p, p'=1, \ldots, N$,
and $N \in \left\lbrace N_{\rm AP},N_{\rm UE}\right \rbrace$.
We thus introduce the notation:
\begin{equation}
	\begin{array}{lll}
		\mathbb{v}_{k,s,j}= \ds \bW^H_{N_{\rm UE}} \bv_{k,s,j}, \quad 
		\mathbb{u}_{m,s,i}= \ds \bW^H_{N_{\rm AP}} \bu_{m,s,i} \\
		\mathbb{H}^{(s)}_{k,m}(q)= \ds \bW^H_{N_{\rm UE}} \mathcal{H}^{(s)}_{k,m}(q) \bW_{N_{\rm AP}}.
	\end{array}
\end{equation}
The vector $\mathbb{u}_{m,s,i}$, to be used at the $m$-th AP in the $s$-th beacon slot and on the $i$-th RF chain, is defined by the data-pattern. More precisely, letting ${\cal A}_d \in \{ 1, 2, \ldots, M\}$ denote the set of APs that have been assigned the $d$-th data-pattern, $\mathbb{u}_{m,s,i}$ follows 
\begin{equation}
	\mathbb{u}_{m,s,i}=\mathbb{d}_{d,s,i}, \; \forall \, m \in \mathcal{A}_d\, ,
	\label{eq:beamformers_resorce}
\end{equation}
i.e., all the APs using the $d$-th data-patterns in the $s$-th beacon slot and on the $i$-th RF chain use the transmit beamformer $\mathbb{d}_{d,s,i}$. 
We will use pseudo-random multi-finger transmit and receive beamformers
\cite{hassanieh2018fast,Caire_scalable_robust_BA_TCOM2018,song2019efficient}.

{In particular, the pseudo-random beamformers transmitted by the APs in the system during the $T$ beacon slots are defined as the collection of sets $\mathcal{C}_{\rm AP}= \lbrace \mathcal{U}_{d,s,i} , d=1, \ldots , \widetilde{D}, \, s=1,\ldots, T, \, i=1,\ldots , n_{\rm AP}\rbrace$, where $\mathcal{U}_{d,s,i}$ is the angle domain support, i.e., the subset of quantized angles in the virtual beam space representation. 
Otherwise stated, we assume $\left| \mathcal{U}_{d,s,i} \right|= \nu_{\rm AP} \leq N_{\rm AP} ,\, \forall \; d,s,i$ and $\mathbb{d}_{d,s,i} =\frac{\mathbf{1}_{\mathcal{U}_{d,s,i} }}{\sqrt{\nu_{\rm AP}}}$, where $\mathbf{1}_{\mathcal{U}_{d,s,i}}$ is the $N_{\rm AP}$-dimensional vector with 1 at (randomly generated) positions  in the support set $\mathcal{U}_{d,s,i}$ and 0 elsewhere. 

While the data-patterns to be used used at the APs are pre-determined and known to all the network entities, the pseudo-random beamforming codebook used at the UEs can be locally customized, i.e., the $k$-th UE can autonomously choose its own combining codebook defined by the collections of sets $\mathcal{C}^{(k)}_{\rm UE}= \lbrace \mathcal{V}_{k,s,j} ,\, s=1,\ldots, T, \, j=1,\ldots , n_{\rm UE}\rbrace$, where $\mathcal{V}_{k,s,j}$ is the angle domain support defining the directions from which the $k$-th UE collects the signal power in the $s$-th beacon slot and on the $j$-th RF chain. 
Again,  we assume $\left| \mathcal{V}_{k,s,j} \right|= \nu_{\rm UE} \leq N_{\rm UE} ,\, \forall \; k,s,j$ and $\mathbb{v}_{k,s,j} =\frac{\mathbf{1}_{\mathcal{V}_{k,s,j} }}{\sqrt{\nu_{\rm UE}}}$, where $\mathbf{1}_{\mathcal{V}_{k,s,j}}$ is the $N_{\rm UE}$-dimensional vector with 1 at the randomly chosen positions in the support set $\mathcal{V}_{k,s,j}$ and 0 elsewhere. }

The main mathematical symbols used in this paper are summarized in Table \ref{Symbols_table}.

\begin{table*}[]
	\caption{Meaning of the main mathematical symbols}
	\label{Symbols_table}
	\begin{center}
\begin{tabulary}{\columnwidth}{ |p{2.2cm}|p{14cm}|}
		\hline
		\textbf{Symbol} & \textbf{Interpretation} \\ \hline
		$\bH_{k,m}^{(s)}(\tau)$ & $(N_{\rm UE} \times N_{\rm AP})$-dimensional matrix-valued impulse response of the channel between the $m$-th AP and the $k$-th UE in the $s$-th beacon slot\\ \hline
		${\cal D}_d$ & $d$-th data-pattern with $d=1,\ldots,D$, which contains for each of the $n_{\rm AP}$ transmit RF chains, the $Q$ subcarriers \textit{and} the beamformers to be used in each beacon slot \\ \hline
		$\bm{\phi}_{\ell}$ & $\ell$-th pilot sequence with $\ell=1,\ldots,S$, of length $S$ to guarantee the orthogonality between $S$ APs that use the same data-pattern  \\ \hline
		$\bu_{m,s,i}$ & the $N_{\rm AP}$-dimensional transmit beamforming vector used by the $m$-th AP on the $i$-th RF chain in the $s$-th beacon slot  \\ \hline
		$\bv_{k,s,j}$ & the $N_{\rm UE}$-dimensional receive beamforming vector used by the $k$-th UE on the $j$-th RF chain in the $s$-th beacon slot \\ \hline
		$\mathcal{H}_{k,m}^{(s)}(q)$ &  matrix-valued Fourier transform of the channel impulse response $\bH_{k,m}^{(s)}(\tau)$ on the $q$-th subcarrier \\ \hline
		$\Theta, \Phi$ & discrete set of AoDs and AoAs \\ \hline
		$\bW_{N_{\rm AP}}$, $\bW_{N_{\rm UE}}$ & $N_{\rm AP}$-dimensional and $N_{\rm UE}$-dimensional unitary discrete Fourier transform (DFT) matrices  \\ \hline
		$\mathbb{u}_{m,s,i}$ & projection of $\bu_{m,s,i}$ on the finite-dimensional (discrete) beamspace representation \\ \hline
		$\mathbb{v}_{k,s,j}$ & projection of $\bv_{k,s,j}$ on the finite-dimensional (discrete) beamspace representation \\ \hline
		$\mathbb{H}^{(s)}_{k,m}(q)$ & projection of $\mathcal{H}^{(s)}_{k,m}(q)$ on the finite-dimensional (discrete) beamspace representation  \\ \hline
		$\mathbb{d}_{d,s,i}$ & projection on the finite-dimensional (discrete) beamspace representation of the transmit beamforming vector of all the APs using the $d$-th data-patterns in the $s$-th beacon slot and on the $i$-th RF chain\\ \hline
	\end{tabulary}
\end{center}
\end{table*}

\section{Signal processing at the UE for beam alignment}
Given the definitions in Section \ref{BA_signal_model}, Eq. \eqref{eq:Yskj} can be shown to be written as
\begin{equation}
	\begin{array}{llll}
		Y_{k,p,j,i}(q)=& \ds \frac{1}{\sqrt{n_{\rm UE}}} \ds  \sum _{m=1}^M \mathbb{v}^H_{k,s(p),j}
		\mathbb{H}^{(s)}_{k,m}(q) \mathbb{u}_{m,s(p),i} \times \\ & \; 
		X_{m,s(p),i}(q)  + Z_{k,p,j,i}(q) \; , \label{eq:Yskj2}
	\end{array}
\end{equation}
{where $Z_{k,p,j,i}(q)$ is the AWGN contribution.}
Since each data-pattern adopts a disjoint set of subcarriers, the UE can operate on $\widetilde{D}$ different sets of observables, isolating the contribution from each AP transmit RF chain. 
The $(d,\ell)$-th set of observables at the $k$-th UE, that we denote by ${\cal O}_k^{(d,\ell)}$, is thus expressed as
\begin{equation*}
\begin{array}{llll}
{\cal O}_k^{(d,\ell)} &=\left\{\overline{Y}^{(d,\ell)}_{k,p,j,i}(q), \; q \in {\cal L}_{d,s,i}, \;  s=1, \ldots, T,  \right. \\ & p=1, \ldots, ST,
\left.  \; i=1, \ldots, n_{\rm AP}, \; j=1, \ldots, n_{\rm UE} \rule{0mm}{7mm} \right\},
\label{eq:finaldata_d}
\end{array}
\end{equation*}
with $d=1,\ldots,D$, $\ell=1,\ldots, S$ and 
\begin{equation}
\begin{array}{llll}
\overline{Y}^{(d,\ell)}_{k,p,j,i}(q)=& \ds \frac{\sqrt{\beta}}{\sqrt{n_{\rm UE}}} \ds  \sum _{m \in \mathcal{A}_d}  \mathbb{v}^H_{k,s(p),j}
\mathbb{H}^{(s)}_{k,m}(q) \mathbb{u}_{m,s(p),i} 
 \times \\ & 
e^{\i \widetilde{\phi}_{\ell(m)}(p \mod s)} + \overline{Z}^{(d,\ell)}_{k,p,j}(q)\, , 
\end{array}
\end{equation}
{with $\overline{Z}^{(d,\ell)}_{k,p,j}(q)$ the AWGN contribution on the $(d,\ell)$-th pair.}
{Based on the above data, the following averaged quadratic observable is built:
\begin{equation}
c_{k,s,j,i}^{(d,\ell)}=\ds\frac{1}{Q S}\!\!\!\sum_{q\in {\cal L}_{d,s,i}}\!\!\left|  \ds \sum_{p=(s-1)S+1}^{sS} \overline{Y}^{(d)}_{k,p,j,i}(q)e^{-\i \widetilde{\phi}_{\ell}(p \!\! \mod \! s)}\right|^2,
\label{eq:averaged}
\end{equation}
for all $d=1,\ldots, D, \, \ell=1,\ldots,S,\, s=1,\ldots,S,\, i=1,\ldots, n_{\rm AP}, \, j=1,\ldots, n_{\rm UE}, \, k=1,\ldots, K$.}

Based on data in \eqref{eq:averaged}, two different algorithms are here proposed in order to extract the information on the AoA and AoD of the strongest path from the closest AP using the $d$-th data-pattern and the $\ell$-th pilot sequence.

\subsection{Processing based on stacked collection of observables (SCO)}
This algorithm is inspired by the one in \cite{Caire_scalable_robust_BA_TCOM2018} for a single-AP system. First of all, at the $k$-th UE the measurements in Eq. \eqref{eq:averaged} are collected for all the values of $i$, $j$ and $s$ and grouped into the following vector:
\begin{equation}
	\mathbf{c}_k^{(d,\ell)}\!=\!\!\left[c_{k,1,1,1}^{(d,\ell)}, \ldots, c_{k,1,n_{\rm UE},n_{\rm AP}}^{(d,\ell)}, c_{k,2,1,1}^{(d,\ell)}, \ldots, 
	c_{k,T,n_{\rm UE},n_{\rm AP}}^{(d,\ell)}\right]^T .  
\end{equation}
Next, let
{ 
\begin{equation}
	\mathbb{b}_{k,s,j,i}^{(d)}=\ds \frac{\mathbb{d}_{d,s,i} \otimes \mathbb{v}_{k,s,j}}{\| \mathbb{d}_{d,s,i} \| \| \mathbb{v}_{k,s,j}\|} \; ,
\end{equation}}
and form the $(n_{\rm AP}n_{\rm UE}T \times N_{\rm AP}N_{\rm UE})$-dimensional matrix 
\begin{equation}
	\begin{array}{ll}
		\mathbf{B}_k^{(d)} = & \left[\mathbb{b}_{k,1,1,1}^{(d)}, \ldots, \mathbb{b}_{k,1,n_{\rm UE},n_{\rm AP}}^{(d)}, \right. \\ & \left.  
		\mathbb{b}_{k,2,1,1}^{(d)}, \ldots, 
		\mathbb{b}_{k,T,n_{\rm UE},n_{\rm AP}}^{(d)}\right]^T \; .
\end{array}\end{equation}

Note that the matrix $\mathbf{B}_k^{(d)}$ depends only on the beamforming vectors used in the $d$-th data-pattern and not on the pilot sequences used by the APs. 

Based on the above notation, the following optimization problem can be considered:
\begin{equation}
	\bm{\xi}_{k}^{(d,\ell)}= \mbox{arg} \, \min_{\mathbf{x}} \left\| \mathbf{B}_k^{(d)} \mathbf{x} + \sigma^2 \mathbf{1}_{n_{\rm AP}n_{\rm UE}T\times 1} - \mathbf{c}_k^{(d,\ell)}\right\|^2 \; .
	\label{eq:problem_kd}
\end{equation}
The solution $\bm{\xi}_{k}^{(d,\ell)}$ to Problem \eqref{eq:problem_kd}  is a $(N_{\rm AP}N_{\rm UE})$-dimensional vector that can be arranged in a $(N_{\rm AP}\times N_{\rm UE})$-dimensional matrix, $\bm{\Xi}_{k}^{(d,\ell)}$ say, where each entry can be associated to a pair (AoD, AoA) associated to a possible propagation path coming from the APs using the $d$-th data-pattern and the $\ell$-th pilot sequence. Each entry of $\bm{\Xi}_{k}^{(d,\ell)}$ contains, for each possible pair (AoD, AoA), an estimate of the channel power; it thus follows that the largest entry of  $\bm{\Xi}_{k}^{(d,\ell)}$ is an indicator of the dominant path between the $k$-th UE and the APs using the $d$-th data-pattern and the $\ell$-th pilot sequence; likewise, the second largest entry of 
$\bm{\Xi}_{k}^{(d,\ell)}$ can be associated to the second strongest path and so on. 
The (convex) optimization problem \eqref{eq:problem_kd} is generally referred to as \textit{Non-Negative Least-Squares} (NNLS), and has been well investigated in the literature \cite{donoho1992maximum,slawski2013non,bruckstein2008uniqueness}. In terms of numerical implementation, the NNLS can be posed as an unconstrained LS problem over the positive orthant and can be solved by several efficient techniques
such as Gradient Projection or Primal-Dual techniques with an affordable computational complexity, generally significantly smaller than compressed sensing techniques\cite{bertsekas2015convex,kim2010tackling}.

\subsection{Processing based on matrix-valued collection of observables (MCO)} 
The measurements in \eqref{eq:averaged} can be collected for all the values of $i$, $j$ and $s$ and grouped in the $(N_{\rm UE}\times N_{\rm AP})$-dimensional matrix $\mathbf{C}_k^{(d,\ell)}$ 
{
\begin{equation}
	\mathbf{C}_k^{(d,\ell)}=\ds \sum_{s=1}^T \widetilde{\mathbf{C}}_k^{(d,\ell)}(s)
	\label{C_matrix}
\end{equation}
where the generic entry of matrix $\widetilde{\mathbf{C}}_k^{(d,\ell)}(s)$ is
\begin{equation}
	\left(\widetilde{\mathbf{C}}_k^{(d,\ell)}(s)\right)_{(h, h')}= I_{h, h'}c_{k,i,j,s}^{(d,\ell)}
	\label{C_tilde_matrix}
\end{equation}}
where $I_{h, h'}$ is one if the $h'$-th transmit direction at the AP and the $h$-th receive direction are active when the measurements leading to  $c_{k,i,j,s}^{(d,\ell)}$ are made. Given the matrix $\mathbf{C}_k^{(d,\ell)}$,  the positions of its largest entry is an indicator of the dominant path between the $k$-th UE and the AP using the $d$-th data-pattern and the $\ell$-th pilot sequence; likewise, the second largest entry can be associated to the second largest path and so on. In the MCO algorithm, no optimization problem is to be solved to obtain the largest entry of matrix $\mathbf{C}_k^{(d,\ell)}$. The complexity of the MCO procedure is thus considerably lower than the SCO procedure.

{\subsection{MCO-based processing in a dynamic scenario}
In presence of user mobility, the BA must be updated periodically in order to find the best beams and APs to serve each UE when it moves. In the following, we show that the MCO-based BA procedure can be generalized by enriching it with a tracking capability to account for a  dynamic scenario. Denote by $T_{\rm BA}=t_0 S T$ the time that is used to perform  BA, and by $T_{\rm D}$ the time during which we use the estimated beams for the data transmission. We assume that the time is divided in slots of duration $T_{\rm BA} + T_{\rm D}$, and that the first $T_{\rm BA}$ seconds of each slot are used to update the BA information for all the UEs (see Fig. 
 \ref{Fig:dynamic_time_evolution}). 
We denote by $r$ the \textit{frame number index} that counting the number of BA procedures repeated every $T_{\rm D}$ seconds. Denote by $T_{\rm C}$, the coherence time of the channel, written as\cite{rappaport2002wireless} 
\begin{equation}
	T_C=\sqrt{\frac{9}{16 \pi f_{D, {\rm max}}^2}} \approx \frac{0.423}{f_{D, {\rm max}}} \, ,
\end{equation}
where $f_{D, {\rm max}}=f_c v_{\rm max}/c$ is the maximum Doppler shift, $f_c$ is the carrier frequency, $v_{\rm max}$ is the maximum speed of the users and $c$ is the speed of light. In practical situations and for typical values of users' speed in urban environments, $T_C<T_{\rm BA}$ and this results in a fast fading component of the channel that \textit{ages} in time, i.e. in different beacon slots during the $r$-th iteration of the BA procedure. This behaviour can be easily included in our model since we assume that the fast fading component of the channel, $\alpha_{k,m,\ell'}^{(s)}$ changes between different beacon slots in the model in Eq. \eqref{eq:channelmodel_km}; see for instance the model in reference \cite{DAndrea_EUSIPCO2021}\footnote{We omit here the details for the sake of brevity.}.
In order to update the BA information, the MCO-based algorithm is used; specifically,
the $r$-th update of the BA is performed using the following matrix:
\begin{equation}
	\mathbf{C}_{k,r}^{(d,\ell)}=\ds \sum_{\widetilde{r}=1}^r \ds \sum_{s=1}^T \lambda_{\rm F}^{rT-(\widetilde{r}-1)T+s}\widetilde{\mathbf{C}}_k^{(d,\ell)}\left[(\widetilde{r}-1)T+s\right] \, .
	\label{C_matrix_r_mobility}
\end{equation}
In \eqref{C_matrix_r_mobility}, the matrix 
 $\widetilde{\mathbf{C}}_k^{(d,\ell)}(s)$ is defined as in Eq. \eqref{C_tilde_matrix}, and represents the collected data during the $s$-th beacon slot, while $\lambda_{\rm F}$ is a close-to-unity forgetting factor used to introduce the BA tracking capability. 
}

\begin{small}\begin{figure*}
	\begin{center}
	\includegraphics[scale=0.4]{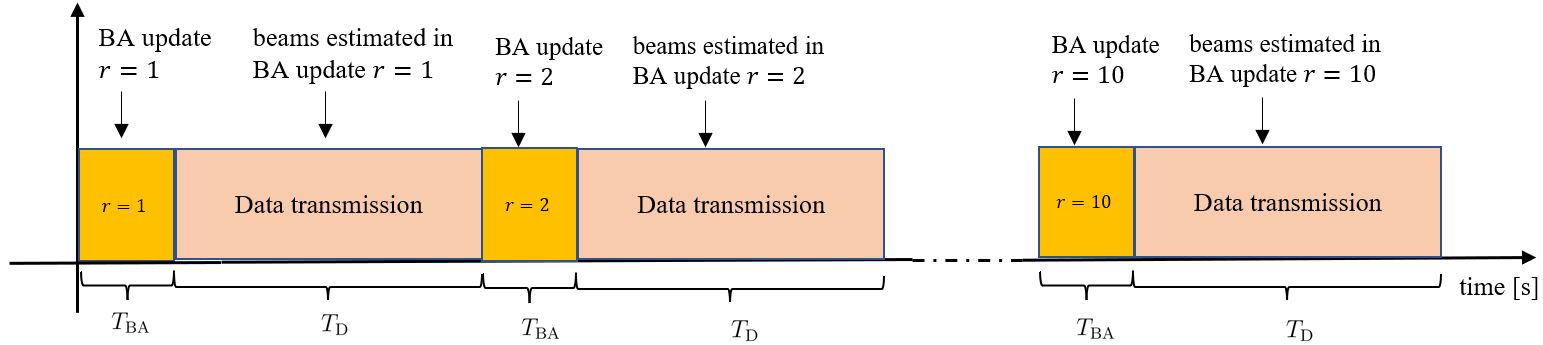}
	\caption{{The temporal evolution of the proposed BA procedure to account for a dynamic environment.}}
	\label{Fig:dynamic_time_evolution}
	\end{center}
\end{figure*}
\end{small}
\section{Data transmission phase} \label{Data_section}

Once the BA procedure detailed in Section \ref{BA_procedure} is over, each UE knows the estimates of the strongest (AoA, AoD) pairs, a strength indicator, the data-pattern and the pilot index on which each of these estimates were obtained. Otherwise stated, if pilot-less data-patterns are used, the $k$-th UE has the following information
\begin{equation}
	\begin{array}{lll}
		{\cal B}_k=&\left\{ \rho^{(k)}_{d}, \left( \widehat{h}_{k,d}, \widehat{h}'_{k,d}\right), \,  \quad d=1, \ldots, D \right\}\, ,
	\end{array}
	\label{eq:Bk_pilot_less}
\end{equation}
where $\rho^{(k)}_{d}$ is the strength indicator of the strongest estimated path at the $k$-th UE on the $d$-th data-pattern, and $\left( \widehat{h}_{k,d}, \widehat{h}'_{(k,d)}\right)$ is the position of this maximum, which is tied to the estimated (AoD, AoA) pair. 
If, instead, the pilot-based  data-patterns are used, the $k$-th UE has the following information
\begin{equation}
	\begin{array}{lll}
		{\cal B}_k=&\left\{ \rho^{(k)}_{d,\ell}, \left( \widehat{h}_{k,d,\ell}, \widehat{h}'_{k,d,\ell}\right), 
		\quad d=1, \ldots, D, \, \ell=1,\ldots,S \right\}\, ,
	\end{array}
	\label{eq:Bk_pilot_based}
\end{equation}
where $\rho^{(k)}_{d,\ell}$ is the strongest path strength indicator estimated at the $k$-th UE on the $d$-th data-pattern and on the $\ell$-th pilot sequence, and, again, $\left( \widehat{h}_{k,d,\ell}, \widehat{h}'_{k,d,\ell}\right)$ describe the position of this maximum. 

Now, association between the APs and the UEs is to be performed. Optimal association is a complicated combinatorial task that is out of the scope of this paper. We will thus use a simple association rule, by assuming that each UE is associated to the 
$N_D$ APs for which the largest strength indicators have been estimated\footnote{
Notice also that, for large number of UEs, the network could also decide to schedule only part of the UEs in the system, in order to avoid unsatisfactory performance levels due to a too much loaded system.}.
More precisely, when the BA procedure is over, each UE announces to the network, using a reliable sub-6GHz feedback channel, 
its position, its ID, and the IDs of the $N_D$ data-patterns corresponding to the $N_D$ largest estimated strength indicators.
The network gathers such information and associates each UE to the $N_D$ closest APs that are using the data-patterns whose IDs have been announced by the UE. The UE-AP resulting association is broadcasted to the UEs using the downlink sub-6 GHz control channel and to the APs using the fronthaul connection with the CPU. Information on the beam to be used is also communicated to the APs.  
In particular, the AP in $\mathcal{A}_{d_{k,n}}$ using the $\ell_{k,n}$-th pilot sequence and \textit{nearest} to the $k$-th UE, uses the $h'_{k,n}$-th column of the matrix $\bW_{N_{\rm AP}}$ to communicate with the $k$-th UE; similarly,  the $k$-th UE uses the $h_{k,n}$-th column of the matrix $\bW_{N_{\rm UE}}$ to communicate with the AP, and this assignment is made for $n=1,\ldots, N_D$. 
Accordingly, if we denote by  $\widehat{\bu}_{k,m}$  the beamforming vector used at the $m$-th AP to communicate with the $k$-th UE and by $\widehat{\bv}_{k,n}$ the $n$-th beamforming vector at the $k$-th UE, we have 
\begin{equation}
	\widehat{\bu}_{k,m}=\left(\bW_{N_{\rm AP}}\right)_{(:, h'_{k,n})}\, ,
	\label{U_beamformers}
\end{equation}
with $m$ is the index of AP using the $d_{k,n}$-th data-pattern and the $\ell_{k,n}$ pilot sequence which is the \textit{nearest} to the $k$-th UE, and 
\begin{equation}
	\widehat{\bv}_{k,n}=\left(\bW_{N_{\rm UE}}\right)_{(:, h_{k,n})}, \, n=1,\ldots, N_D.
	\label{V_beamformers}
\end{equation}
For future reference, we also introduce the binary-valued association variable 
$a_{k,m}$, which is $1$ if the $k$-th UE is served by the $m$-th AP and $0$ otherwise.

\subsection{Downlink data transmission}
On the downlink, the signal transmitted by the $m$-th AP on the $q$-th subcarrier is the following $N_{\rm AP}$-dimensional vector
\begin{equation}
	\mathbf{s}_m(q) = \ds \sum_{j=1}^K \ds a_{j,m} \sqrt{\eta_{j,m}^{\rm DL}} \widehat{\bu}_{j,m} s_j^{\rm DL}(q) \; ,
	\label{eq:transmittedscalar_a}
\end{equation}
where $\eta_{k,m}^{\rm DL}$ is a scalar coefficient controlling the power transmitted by the $m$-th AP to the $k$-th UE, and $s_k^{\rm DL}(q)$ is the  unit-energy  data symbol to be sent to the $k$-th UE on the $q$-th subcarrier.

Letting $P^{{\rm DL}}_m$ denote the overall transmitted power by the $m$-th AP, the normalized transmit power must satisfy the constraint
\begin{equation}
	\norm{\mathbf{s}_m(q)}^2=\ds \sum_{j=1}^K {\eta_{k,m}^{\rm DL}} \leq P^{{\rm DL}}_m \; .
\end{equation}

Subsequently, each UE receives contributions from all the APs. 
In particular, the $k$-th UE receives on the $q$-th subcarrier the $N_{\rm UE}$-dimensional signal
\begin{equation}
	\mathbf{r}_k(q)=  \ds \sum_{m =1}^M
	\mathcal{H}_{k,m}(q) \mathbf{s}_m(q) + \mathbf{z}_k(q)  ,
	\label{eq:received_data_UE}
\end{equation}
with $\mathbf{z}_k(q)$ being the $N_{\rm UE}$-dimensional additive white Gaussian noise (AWGN) with entries ${\cal CN}(0, \sigma^2_z)$.

{In order to perform the soft estimate for the data symbol, the $k$-th UE uses combiners $\widehat{\bv}_{k,n}, \; n=1,\ldots, N_D$, and the estimate of the transmitted data can be written as
\begin{equation}
	\begin{array}{llll}
		\widehat{s}_k^{\rm DL}(q) \!\!\!\!\!\!& = \ds  \sum_{n=1}^{N_D} \widehat{\bv}_{k,n}^H \mathbf{r}_k(q)
		\\ & 
		= \left( \ds \sum_{m=1}^M a_{k,m} \ds \sqrt{\eta_{k,m}^{\rm DL}} \ds  \sum_{n=1}^{N_D} \widehat{\bv}_{k,n}^H  \mathcal{H}_{k,m}(q)  \widehat{\bu}_{k,m}\right) s_k^{\rm DL}(q) \\ & + 
		\ds \sum_{\substack{j=1 \\ j\neq k}}^K \!\!\left(\!\! \ds \sum_{m=1}^M a_{j,m} \ds \sqrt{\eta_{j,m}^{\rm DL}} \ds  \sum_{n=1}^{N_D} \widehat{\bv}_{k,n}^H  \mathcal{H}_{k,m}(q)  \widehat{\bu}_{j,m}\!\!\right) \!\!s_j^{\rm DL}(q)  \\ & 
		+  
		\ds \sum_{n=1}^{N_D} \widehat{\bv}_{k,n}^H \mathbf{z}_k(q)\; .
	\end{array}
	\label{eq:received_data_UE2}
\end{equation}
}
Based on \eqref{eq:received_data_UE2}, it is straightforward to express the downlink signal to interference plus noise ratio (SINR)  of the $k$-th UE on the $q$-th subcarrier as reported in \eqref{SINR_DL} at the top of next page.
\begin{figure*}
	\begin{equation}
		\text{SINR}_{k,q}^{\rm DL}= \frac{\ds \left| \ds \sum_{m=1}^M a_{k,m} \ds \sqrt{\eta_{k,m}^{\rm DL}} \ds  \sum_{n=1}^{N_D} \widehat{\bv}_{k,n}^H  \mathcal{H}_{k,m}(q)  \widehat{\bu}_{k,m}\right|^2}{\ds \sum_{\substack{j=1 \\ j\neq k}}^K \left| \ds \sum_{m=1}^M a_{j,m} \ds \sqrt{\eta_{j,m}^{\rm DL}} \ds  \sum_{n=1}^{N_D} \widehat{\bv}_{k,n}^H  \mathcal{H}_{k,m}(q)  \widehat{\bu}_{j,m} \right|^2+\sigma^2_z \ds \sum_{n=1}^{N_D} \widehat{\bv}_{k,n}^H \widehat{\bv}_{k,n} } 
		\label{SINR_DL}
	\end{equation}
\end{figure*}

\subsection{Uplink data transmission}
In the uplink, UEs send their data symbols  using the beamforming vectors $\widehat{\bv}_{k,n}, \; n=1,\ldots, N_D$ in Eq. \eqref{V_beamformers}. Accordingly,  the signal transmitted on the uplink by the $k$-th UE on the $q$-th subcarrier is
\begin{equation}
	{\mathbf{t}}_k(q)=\ds \sum_{n=1}^{N_D} \ds \sqrt{\eta_{k,n}^{\rm UL}} \widehat{\bv}_{k,n} {s}^{\rm UL}_k(q)\; ,
\end{equation}
with ${\eta_{k,n}^{\rm UL}}$ the uplink transmit power of the $k$-th UE in the $n$-th direction and ${s}^{\rm UL}_k(q)$ the uplink data symbol of the $k$-th UE. Letting $P^{{\rm UL}}_k$ denote the overall transmitted power by the $k$-th AP, the normalized transmit power must satisfy the constraint
$
	\norm{\mathbf{t}_k(q)}^2=\ds \sum_{n=1}^{N_D} {\eta_{k,n}^{\rm UL}} \leq P^{{\rm UL}}_k $.

As a result, the $N_{\rm AP}$-dimensional signal $\mathbf{y}_m(q)$ received at the $m$-th AP on the $q$-th subcarrier can be expressed as 
\begin{equation}
	\mathbf{y}_m(q)=\ds \sum_{j=1}^K \mathcal{H}_{j,m}(q)^H {\mathbf{t}}_k(q) +\mathbf{w}_m (q) \; ,
\end{equation}
with $\mathbf{w}_m$ the AWGN vector with entries ${\cal CN}(0, \sigma^2_w)$.

Subsequently, the $m$-th AP which communicates with the $k$-th UE forms its local statistic
$	y_{m,k}(q)= \widehat{\bu}_{k,m}^H \mathbf{y}_m(q)$, 
to be sent to the CPU for uplink data decoding. 
It is easy to show\footnote{Details are omitted for the sake of brevity.} that the resulting 
uplink SINR  for the $k$-th UE on the $q$-th subcarrier can be written as in \eqref{SINR_UL} at the top of next page.
\begin{figure*}
	\begin{equation}
		\text{SINR}_{k,q}^{\rm UL}= \frac{\ds \left| \ds \sum_{m=1}^M a_{k,m} \ds  \sum_{n=1}^{N_D} \ds \sqrt{\eta_{k,n}^{\rm UL}}  \widehat{\bu}_{k,m}^H  \mathcal{H}_{k,m}(q)^H  \widehat{\bv}_{k,n}\right|^2}{\ds \sum_{\substack{j=1 \\ j\neq k}}^K \left| \ds \sum_{m=1}^M a_{k,m} \ds  \sum_{n=1}^{N_D} \ds \sqrt{\eta_{j,n}^{\rm UL}}  \widehat{\bu}_{k,m}^H  \mathcal{H}_{j,m}(q)^H  \widehat{\bv}_{j,n} \right|^2+\sigma^2_w \ds \sum_{m=1}^{M} a_{k,m} \widehat{\bu}_{k,m}^H \widehat{\bu}_{k,m} } 
		\label{SINR_UL}
	\end{equation}
\begin{center}
	\rule{15cm}{0.2mm}
\end{center}
\end{figure*}

\section{Numerical Results} \label{Numerical_results}
\subsection{Simulation setup}
In our simulation setup, we assume 
a communication bandwidth $W = 123$ MHz centered over the carrier frequency $f_0=28$ GHz. The OFDM subcarrier spacing is 480 kHz and assuming that the length of the cyclic prefix is 7\% of the OFDM symbol duration, i.e., $\tau_{\rm CP}\Delta_f=0.07$, we obtain $t_0=2.23 \mu$s and $N_C=256$ subcarriers. A beacon slot is assumed to contain $S=8$ OFDM symbols.
The antenna height at the APs is $10$ m, while at the UEs it is $1.65$ m. The additive thermal noise is assumed to have a power spectral density of $-174$ dBm/Hz, while the front-end receiver at the APs and at the MSs is assumed to have a noise figure of $9$ dB. We consider a square area of 400m $\times$ 400 m, with $M=50$ APs and $K=10$ UEs; the APs and UEs are equipped with ULAs of $N_{\rm AP}=32$ and $N_{\rm UE}=16$ antennas, respectively, the number of RF chains at the APs and UEs are $n_{\rm AP}=8$ and $n_{\rm UE}=4$, respectively. We assume a number of total scatterers, $N_{\rm s}=5000$ say, common to all the APs and UEs and uniformly distributed in the simulation area. In order to model the signal blockage, we assume that the communication between the $m$-th AP and the $k$-th UE takes place via the $n$-th scatterer, i.e., the $n$-th scatterer is one of the effective $L_{k,m}$ contributing in the channel in Eq. \eqref{eq:channelmodel_km}, if the rays between the $m$-th AP and the $n$-th scatterer and the $k$-th UE and the $n$-th scatterer \emph{simultaneously} exist.
We assume a link exists between two entities, in our case one AP/UE and one scatterer, if they are in LoS, with a probability, $P_{\rm LOS}(d)$ depending on the distance between the two entities, $d$ say. 
For $P_{\rm LOS}(d)$ we use the model in \cite{ghosh20165g_WP,5G3PPPlikemodel}:
\begin{equation}
	P_{\rm LOS}(d)=\text{min}\left(\frac{20}{d},1\right)\left(1-e^{-\frac{d}{39}}\right)+e^{-\frac{d}{39}} \; ,
\end{equation} 
with $d$ the distance between the AP/UE and the intended scatterer.
For the channel model in Eq. \eqref{eq:channelmodel_km} the variance of the complex gain associated to the $\ell'$-th path between the $k$-th UE and the $m$-th AP, $\gamma_{k,m,\ell'}$ is obtained as \cite{ghosh20165g_WP} 
\begin{equation}
	\gamma_{k,m,\ell'}\! =\! - 20\log_{10}\!\!\left(\!\frac{4\pi}{\lambda}\!\right) \\ - 10\wt{n}\! \log_{10}\left(r_{k,m,\ell'}\right) - X_{\rm SF} \; ,
\end{equation}
where $r_{k,m,\ell'}$ is the length of the path, $\wt{n}$ is the path loss exponent, $X_{\rm SF}$ is  the zero-mean, $\sigma_X^2$-variance Gaussian-distributed shadow fading term in logarithmic units, and $\lambda$ is the wavelength. We use the parameters of the Urban Microcellular (UMi) Street-Canyon environment, i.e., $n=3.19 \; , \sigma_X=8.2 \; \text{dB}$\cite{ghosh20165g_WP}.  The propagation delay associated with the $\ell'$-th path between the $k$-th UE and the $m$-th AP is written as $\tau_{k,m,\ell'}=r_{k,m,\ell'}/c$, with $c$ the speed of light.
 The total power transmitted by the APs over all the subcarriers during the BA phase is denoted as $P_{\rm BA}$, and consequently $\beta=P_{\rm BA}/N_C$. In the following, we assume $P_{\rm BA}=10$ dBW. For the data transmission phase, we assume equal stream power allocation both in uplink and downlink; in particular, denoting by $P_m^{\rm DL}$ and $P_k^{\rm UL}$ the available power at the $m$-th AP and at the $k$-th UE, the downlink and uplink power control coefficients are expressed as
 \[
 	\eta_{k,m}^{\rm DL}=\left \lbrace
 	\begin{array}{lllll}
 		&\ds \frac{P_m^{\rm DL}}{\ds \sum_{j=1}^K a_{j,m}} \, &\text{if} \; a_{k,m}=1,
 		\\
 		& 0 & \, \text{otherwise,} 
 	\end{array}\right.\]
 and $\eta_{k,n}^{\rm UL}=\ds \frac{P_k^{\rm UL}}{N_D}$, $n=1,\ldots, N_D$,
 respectively. In the simulations, we used $P_m^{\rm DL}=10$ dBW and $P_k^{\rm UL}=3$ dBW.

\begin{figure*}
	\begin{center}
		\includegraphics[scale=0.5]{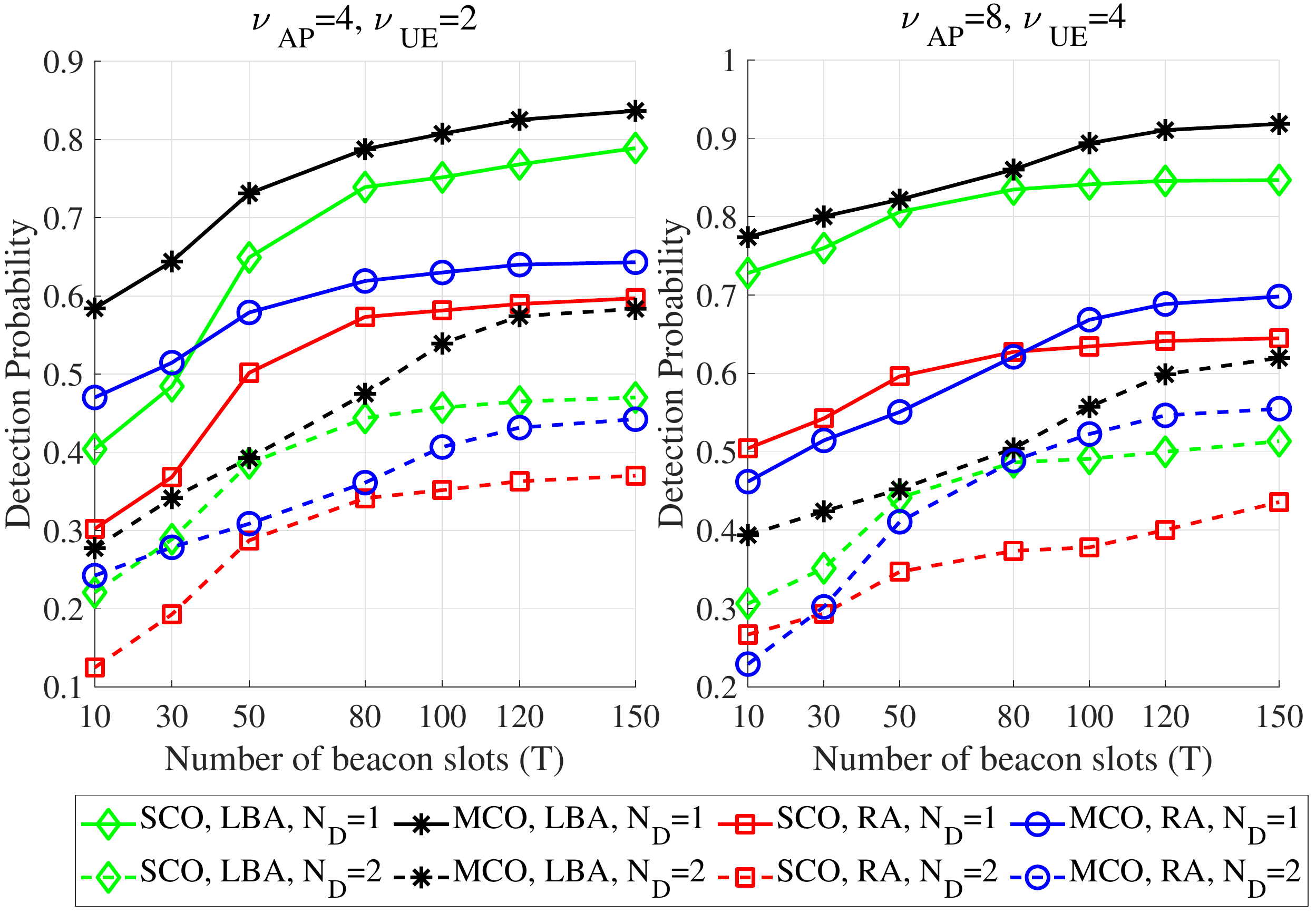}
	\end{center}
	\caption{{\small Detection probability of the pilot-less SCO and MCO proposed schemes versus the number of beacon slots $T$ for different values of $N_D$, $\nu_{\rm AP}$, $\nu_{\rm UE}$ in the cases of location-based assignment (LBA) and random assignment (RA) of the data-patterns. Parameters: $M=50, K=10, N_{\rm AP}=32$, $N_{\rm UE}=16$, $n_{\rm AP}=8$, $n_{\rm UE}=4$, $D=8$.}}
	\label{Fig_Detection_prob_pilot_less_D8}
\end{figure*}

\begin{figure*}
	\begin{center}
		\includegraphics[scale=0.5]{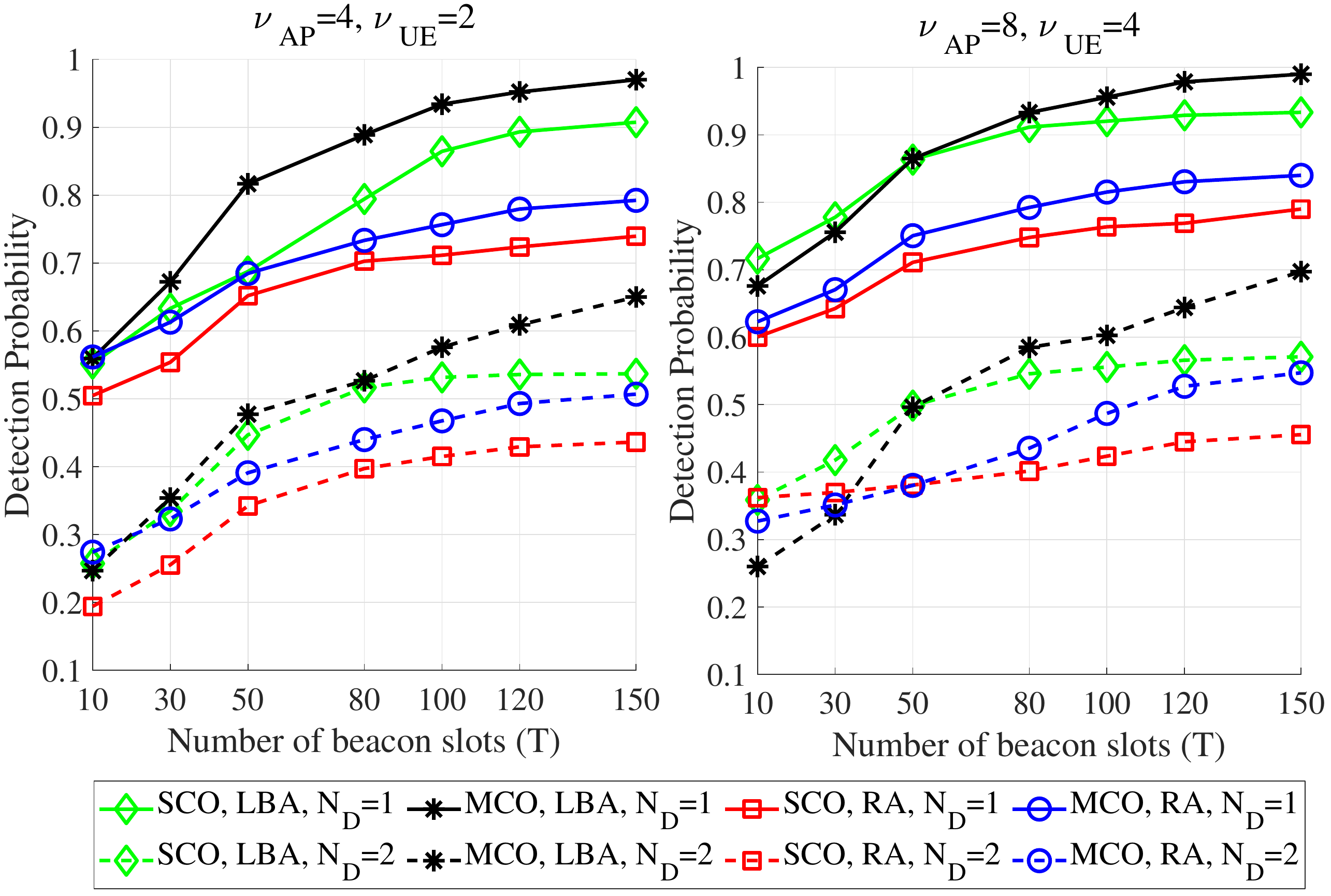}
	\end{center}
	\caption{{\small Detection probability of the pilot-less SCO and MCO proposed schemes versus the number of beacon slots $T$ for different values of $N_D$, $\nu_{\rm AP}$, $\nu_{\rm UE}$ in the cases of location-based assignment (LBA) and random assignment (RA) of the data-patterns. Parameters: $M=50, K=10, N_{\rm AP}=32$, $N_{\rm UE}=16$, $n_{\rm AP}=8$, $n_{\rm UE}=4$, $D=16$.}}
	\label{Fig_Detection_prob_pilot_less_D16}
\end{figure*}

\subsection{Results and comments}

The considered performance measure is the probability of correct detection at the UE of the AoD and AoA of the $N_D$ strongest paths (one for each AP), i.e. the probability that a UE detects the correct AoD and AoA for the strongest path from the $N_D$ best APs\footnote{{This is also referred to as probability of perfect beam alignment.}}.
%
Figs. \ref{Fig_Detection_prob_pilot_less_D8} and \ref{Fig_Detection_prob_pilot_less_D16} show such detection probability versus the number of  beacon slots $T$ used for the BA procedure. In Fig. \ref{Fig_Detection_prob_pilot_less_D8}, there are $D=8$ different data-patterns, while in Fig. \ref{Fig_Detection_prob_pilot_less_D16} the number of data-patterns is $D=16$. Each figure shows the performance for two different pairs of the parameters $(\nu_{\rm AP}, \nu_{\rm UE})$, i.e. the number of active fingers in the beamformers used at the APs and at the UEs, respectively. The figures refer to the case of pilot-less  data-patterns. In order to show the merits of the proposed location-based data-pattern assignment procedure, detailed in Section \ref{LBA_procedure}, we also report the performance corresponding to the case in which a random assignment (RA) of the data-patterns to the APs is performed. Inspecting the figure, it is seen that for all the considered cases the detection probability increases with the number of beacon slots used for the BA phase, which confirms the validity of the proposed approach.
Regarding the comparison between the MCO and SCO procedures for BA, 
we can see that the MCO, albeit being simpler, achieves much better performance than the SCO; moreover, results show that  see that the increase in the parameter $D$ improves the detection capability of the system; however $D$ cannot be increased too much since this corresponds to a smaller value of $Q$, the number of carriers assigned to each AP RF chain. Larger values for the parameters $\nu_{\rm AP}$ and $\nu_{\rm UE}$ also bring some performance improvement in the case of low values of $T$. {Finally, the detection probability for $N_D=1$ is obviously larger than that for $N_D=2$, since in the latter situation two paths rather than one are to be correctly detected, which is more challenging. This challenge, for low values of $T$, is seen to be better coped with by the SCO-based algorithm, presumably since such algorithm is inspired by compressed sensing concepts and thus has good performance in the presence of few observations.}

\begin{figure*}
	\begin{center}
		\includegraphics[scale=0.45]{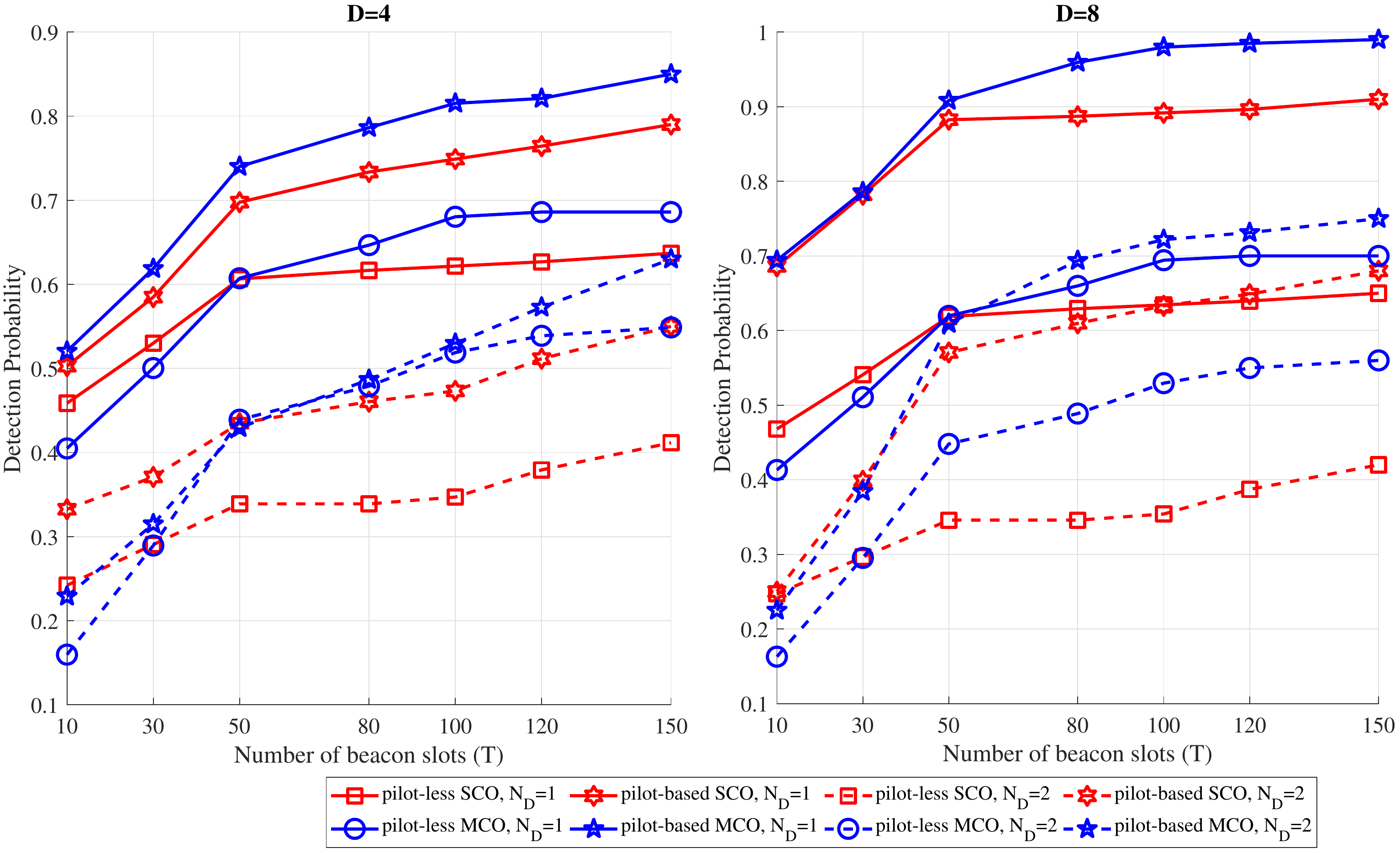}
	\end{center}
	\caption{{\small Detection probability of the proposed pilot-based BA procedure compared with the pilot-less BA procedure versus the number of beacon slots $T$ for different values of $N_D$ and $D$. Parameters: $M=50, K=10, N_{\rm AP}=32$, $N_{\rm UE}=16$, $\nu_{\rm AP}=8$, $\nu_{\rm UE}=4$, $n_{\rm AP}=8$, and $n_{\rm UE}=4$, random assignment (RA) of the data-patterns.}}
	\label{Fig_Detection_prob_pilot_based}
\end{figure*}

Fig. \ref{Fig_Detection_prob_pilot_based} provides a comparison between pilot-less and pilot-based data-patterns. It shows the detection probability versus the number of used beacon slots $T$, for two values of $D$, number of different data-patterns. A random assignment (RA) of the data-patterns to the APs is considered here, while the number of active fingers in the beamformers is $\nu_{\rm AP}=8$ and $\nu_{\rm AP}=4$. The pilot sequences in the pilot-based BA procedure are Hadamard sequences with length $S$. 
Clearly, when pilot-based data-patterns are used the performance is much better than when pilot-less data-patterns are employed. 
Focusing on the pilot-based BA procedure, we can also note that in the case of $D=4$, the orthogonality between the APs during the BA procedure is not preserved and so there is a performance degradation compared with the case $D=8$. 

Overall, the shown results prove that the proposed procedures are effective and permit realizing BA in multi-AP multi-UE environments with good performance. The introduction of the orthogonal pilot sequences helps to further increase the detection capability performance of the algorithms.

\begin{figure*}
	\begin{center}
		\includegraphics[scale=0.36]{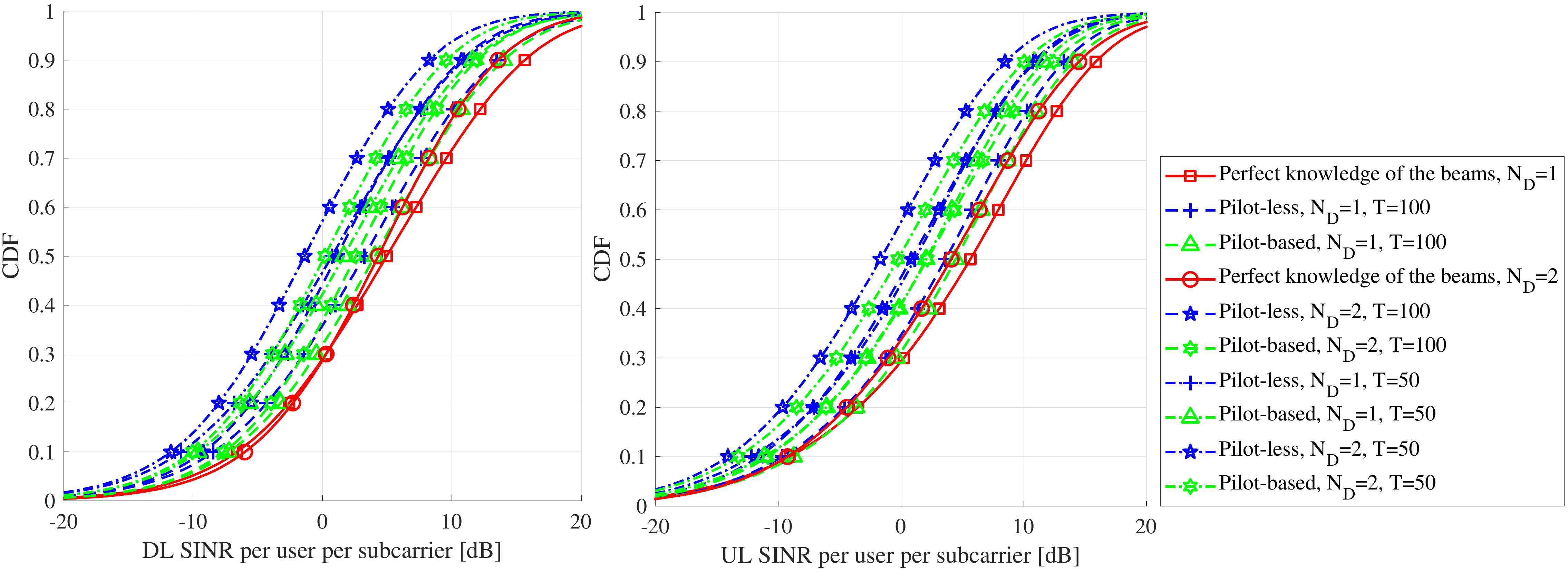}
	\end{center}
	\caption{{\small Cumulative distribution function (CDF) of the downlink (DL) and uplink (UL) SINR per user on each subcarrier of the MCO proposed BA in the cases of pilot-less and pilot-based definition of the data-patterns compared with the case of perfect knowledge of the $N_D$ strongest beams. Two values of the parameter $T$ are reported. Parameters: $M=50, K=10, N_{\rm AP}=32$, $N_{\rm UE}=16$, $n_{\rm AP}=n_{\rm UE}=4$, location-based assignment (LBA) of data-patterns and $D=8$.}}
	\label{Fig_SINR}
\end{figure*}

We now consider the performance during the data transmission phase.
Fig. \ref{Fig_SINR} reports the empirical CDF  of the SINR per user on each subcarrier, evaluated as in Section \ref{Data_section}, for the downlink and the uplink. We compare the performance obtained with the proposed BA procedure based on the MCO technique, with the case of perfect knowledge of the directions of strongest paths. The beamformers at the APs and UEs are thus obtained following Eqs. \eqref{U_beamformers} and \eqref{V_beamformers}, respectively. We can see that, especially in the case of pilot-based  data-patterns, the BA procedure is very effective in finding the strongest beams and these beams can be efficiently used both for the uplink and downlink communication. The obtained SINRs is just few dBs far from the one corresponding to the ideal case of known channel. This is a further confirmation of the effectiveness of the BA procedure here proposed.

\begin{figure}
	\begin{center}
		\includegraphics[scale=0.42]{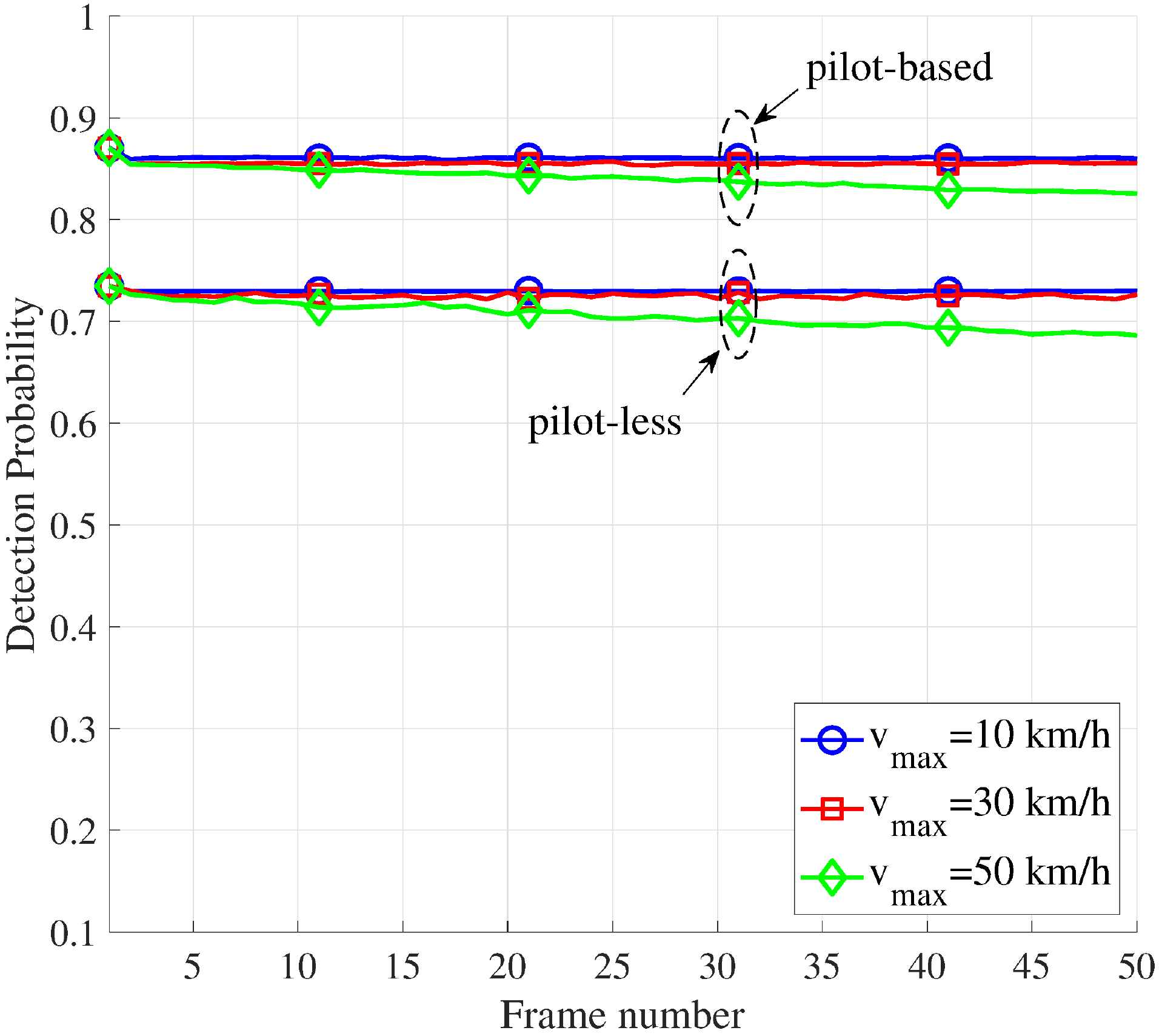}
	\end{center}
	\caption{{\small Detection probability of the proposed MCO procedure in presence of user mobility versus the number of frame number, $r$. Thre values of maximum users' speeds are reported. Parameters: $M=50, K=10, N_{\rm AP}=32$, $N_{\rm UE}=16$, $n_{\rm AP}=n_{\rm UE}=4$, location-based assignment(LBA) of data-patterns, $D=4$, and $\lambda_{\rm F}=0.95$.}}
	\label{Fig_mobility}
\end{figure}

{Finally, we report the performance of the BA procedure in a dynamic scenario. Fig. \ref{Fig_mobility} shows the detection probability versus the frame number, i.e., $r$, assuming three different values for $v_{\rm max}$,  the maximum speed of the users in the simulated area. We assume a common value of $T_{\rm BA}=50$ ms. In the considered scenario, 
 $D=4$ is assumed with LBA assignment of the data-patterns; both pilot-less and pilot-based data-patterns are considered. It can be seen that the detection probability is approximately constant as time evolves, thus implying that the proposed procedure is able to effectively track the beams. }

\section{Conclusions}
This paper has considered the problem of performing BA in a CF-mMIMO network operating at mmWave frequencies. The proposed BA procedure amounts to a protocol involving the CPU, the UEs, the APs and a macro-BS managing a control channel at sub-6 GHz frequency. It enables simultaneous BA of each UE with the strongest beams coming from a pre-defined number of strongest APs. A procedure to assign the data-patterns across the APs has also been proposed. Two different algorithms, to be run at the UE, have been proposed. Of these, the MCO has been shown to achieve better performance  with smaller complexity than the other proposed algorithm, the SCO one. Numerical results have confirmed the effectiveness of the proposed approach both in terms of detection probability and in terms of UL and DL SINR, confirming that BA can be performed in a shared frequency band with a simultaneous operation of several APs and several UEs.

\bibliography{references}
\bibliographystyle{IEEEtran}

\end{document}